\documentclass[superscriptaddress, twocolumn,showpacs,
amssymb,amsmath,nobibnotes,aps,
nofootinbib]{revtex4-1}
\pdfoutput=1
\usepackage{graphicx,subfigure,bm,color,psfrag,hyperref}
\usepackage{amsfonts}
\usepackage{lipsum}
\usepackage{mathtools}
\usepackage{verbatim}
\usepackage[normalem]{ulem}
\usepackage[dvipsnames]{xcolor}
\hypersetup{colorlinks,linkcolor={blue},citecolor={red},urlcolor={cyan}}   

\begin{document}

\title{Accelerating scaling solutions from dark matter particle creation}

\author{Sudip Halder}
\email{sudip.rs@presiuniv.ac.in}
\affiliation{Department of Mathematics, Presidency University, 86/1 College 
Street, Kolkata 700073, India}

\author{Jaume  de Haro}
\email{jaime.haro@upc.edu}
\affiliation{Departament de Matem\`atiques, Universitat Polit\`ecnica de 
Catalunya, Diagonal 647, 08028 Barcelona, Spain}

\author{Supriya Pan}
\email{supriya.maths@presiuniv.ac.in}
\affiliation{Department of Mathematics, Presidency University, 86/1 College 
Street, Kolkata 700073, India}
\affiliation{Institute of Systems Science, Durban University of Technology, 
Durban 4000, Republic of South Africa}

\author{Emmanuel N. Saridakis}
\email{msaridak@noa.gr}
\affiliation {National Observatory of Athens, Lofos Nymfon, 11852 Athens,
Greece}
\affiliation {Departamento de Matem\'{a}ticas, Universidad Cat\'{o}lica del
Norte, Avda. Angamos 0610, Casilla 1280 Antofagasta, Chile}
\affiliation {CAS Key Laboratory for Researches in Galaxies and Cosmology,
Department of Astronomy, University of Science and Technology of China, Hefei,
Anhui 230026, P.R. China}

\author{Tapan Saha}
\email{tapan.maths@presiuniv.ac.in}
\affiliation{Department of Mathematics, Presidency University, 86/1 College 
Street, Kolkata 700073, India}

\author{Subenoy Chakraborty}
\email{schakraborty.math@gmail.com}
\affiliation{Department of Mathematics, Brainware University, Barasat, West 
Bengal 700125, India}

\begin{abstract}
This article opens new window to obtain accelerating scaling attractors without any need of dark energy. We study cosmological dynamics in a two-fluid system where pressureless dark 
matter (DM) undergoes adiabatic particle creation and exchanges energy with a 
barotropic fluid.  Considering six widely used interaction prescriptions, we 
formulate the corresponding autonomous systems in a compact phase space and 
perform a unified dynamical analysis.
We find that  accelerating scaling attractors, namely late-time states where 
both 
fluids coexist with fixed energy fractions, arise only when the interaction is 
controlled by the DM density and energy flows from DM to the second fluid.  
Such attractors appear in the global and local DM-based interactions, and in 
the global mixed case, but are entirely absent when the interaction depends on 
the second fluid or on local mixed terms, which instead drive the universe to a 
DM-dominated accelerating phase. 
These results clarify the unique conditions under which matter creation can 
mimic dark-energy-like behaviour without introducing a dark-energy component.
\end{abstract}

\maketitle

\section{Introduction}
\label{sec-introduction}

One of the major breakthroughs in modern cosmology is the discovery that the
universe is undergoing a phase of late-time accelerated expansion
\cite{SupernovaSearchTeam:1998fmf, SupernovaCosmologyProject:1998vns,
Planck:2018vyg}. Within the framework of General Relativity (GR), standard
cosmic fluids cannot generate such an accelerated phase, thus motivating the
introduction of new ingredients or modifications of the gravitational sector.
Two main directions have emerged, namely the addition of a fluid with large
negative pressure - dark energy (DE) \cite{Copeland:2006wr,Bamba:2012cp}, and a 
broad family
of modified gravity (MG) theories
\cite{CANTATA:2021asi,DeFelice:2010aj, 
Clifton:2011jh,Capozziello:2011et,Cai:2015emx,Nojiri:2017ncd,Bahamonde:2021gfp}
. Among the numerous models in these categories, the
$\Lambda$CDM scenario, in which a cosmological constant is inserted into the
Einstein equations, remains the most economical description and is consistent
with the majority of current observations, including Planck 2018
\cite{Planck:2018vyg}.

Despite its empirical success, $\Lambda$CDM is challenged by well-known
theoretical issues, such as the cosmological constant
\cite{Weinberg:1988cp} and cosmic coincidence
\cite{Zlatev:1998tr} problems, as well as by a set of observational anomalies,
collectively referred to as the ``cosmological tensions''
\cite{DiValentino:2021izs}. These challenges have revived efforts to explore
alternatives beyond $\Lambda$CDM. At present, however, no non-$\Lambda$CDM
model provides a fully satisfactory fit to all available datasets while
remaining free of theoretical or phenomenological difficulties, and 
consequently,
all viable alternatives remain of interest until they are decisively ruled out.

One important class of modified cosmological scenarios arises if one allows for 
mutual interactions between dark energy and dark matter 
sectors \cite{Bolotin:2013jpa,Wang:2016lxa} (also see 
\cite{Chimento:2003iea,Barrow:2006hia,Amendola:2006dg,delCampo:2008jx,
He:2008tn, Basilakos:2008ae,
Gavela:2009cy,
Jamil:2009eb,Lip:2010dr,Chen:2011cy,  Yang:2014gza, 
Faraoni:2014vra,Salvatelli:2014zta,Yang:2014hea,Pan:2012ki, Nunes:2016dlj,
vandeBruck:2016jgg, 
Mukherjee:2016shl,vandeBruck:2016hpz,Cai:2017yww,Yang:2017zjs,
Pan:2017ent, Mifsud:2017fsy, Yang:2018pej,
Yang:2018xlt,Yang:2018qec, Pan:2019jqh, 
Li:2019loh, 
Yang:2019bpr,Yang:2019vni,Paliathanasis:2019hbi,DiValentino:2019ffd,
Gomez-Valent:2020mqn,Paliathanasis:2020wjl,Khyllep:2021wjd,Chatzidakis:2022mpf,
Zhai:2023yny,Li:2023gtu,Li:2023fdk,Forconi:2023hsj,Teixeira:2023zjt,
Paliathanasis:2023moe,Giare:2024ytc,
Li:2024qso,Li:2025owk,Zhai:2025hfi,
vanderWesthuizen:2025iam,Li:2025muv,Li:2025ula,vanderWesthuizen:2025vcb,
vanderWesthuizen:2025mnw, 
Zhang:2025dwu,Figueruelo:2026eis,Paliathanasis:2026ymi}), since in 
this way one can 
alleviate the coincidence problem
\cite{Amendola:1999qq,Amendola:2000uh,
Zimdahl:2001ar,Wetterich:1994bg,Amendola:1999er, 
Cai:2004dk, delCampo:2008sr}, the   $H_0$ tension
\cite{Kumar:2017dnp,DiValentino:2017iww, Yang:2018euj, 
Yang:2018uae,Pan:2019gop,Pan:2020bur,Giare:2024smz},    
 and moreover one can obtain  
accelerating scaling 
solutions~\cite{Billyard:2000bh,Tocchini-Valentini:2001wmi,
Gumjudpai:2005ry,Liu:2005wga,Amendola:2006qi,Chen:2008ft,Boehmer:2008av,
Chen:2008pz, Farajollahi:2011jr, Gomes:2013ema,
Hossain:2014xha,Zhang:2014zfa,Shahalam:2015sja, Halder:2024uao,
Halder:2024aan,Halder:2025ytq}.

On the other hand,  gravitationally induced adiabatic particle production
offers a particularly intriguing mechanism for late-time cosmic acceleration.
In this approach, the expansion of the universe, described by GR, is influenced
by a negative ``creation pressure'' associated with the production of matter
particles. Interestingly, this mechanism is capable of driving accelerated
expansion without invoking any dark energy component or modifications of
gravity. Although several studies have explored this possibility
\cite{Lima:2011hq, Jesus:2011ek, Lima:2012cm, Lima:2014qpa, Lima:2014hda,
Chakraborty:2014fia, Baranov:2015eha, deHaro:2015hdp, Pan:2016jli,
Paliathanasis:2016dhu, Pan:2018ibu}, matter-creation cosmology has not yet
received attention comparable to DE or MG models.

In this work we investigate a generalized interacting matter-creation scenario
consisting of two fluids: pressureless dark matter (DM), endowed with a
matter-creation rate, and a conventional perfect fluid with a barotropic
equation of state. We further allow the two fluids to exchange energy through
a phenomenological interaction term. Introducing interactions in this context
serves two purposes: it broadens the dynamical possibilities of the system,
and it allows us to explore whether accelerating scaling solutions, well known
from interacting DE-DM models \cite{Amendola:1999er}, can emerge in
matter-creation cosmology as well. This question is nontrivial, since without
interaction such solutions do not arise unless the second fluid possesses a
negative equation of state \cite{Halder:2025eze}. As accelerating scaling
solutions are of particular interest for alleviating coincidence-type
problems, it is natural to examine under which interaction prescriptions they
can be realized in two-fluid matter-creation models.

The paper is structured as follows. In Sec.~\ref{sec-basic-eqns} we introduce
the gravitational field equations for interacting matter creation and present
the six interaction prescriptions examined in this work.  In
Sec.~\ref{sec-dyn-analysis} we construct the corresponding autonomous systems,
perform a detailed phase-space analysis for each interaction model, and discuss
their cosmological implications.  Sec.~\ref{sec-comparison} provides a unified
comparison of all models, identifying the general conditions under which
accelerating scaling attractors can arise.  Finally, in
Sec.~\ref{sec-summary} we summarize the main results and outline possible
extensions of the present framework.

\section{Interacting Matter-Creating Cosmologies}
\label{sec-basic-eqns}

We consider a spatially flat Friedmann-Lema\^{i}tre-Robertson-Walker (FLRW)
line element,
\begin{equation}
ds^2 = -dt^2 + a^2(t)\,(dx^2 + dy^2 + dz^2),
\end{equation}
where $a(t)$ is the expansion scale factor. The gravitational sector is
described by Einstein's General Relativity (GR), while the matter sector
consists of two interacting perfect fluids: pressureless dark matter (DM),
  with gravitationally induced adiabatic matter creation, and a
non-matter-creating fluid with a barotropic equation of state. The Friedmann
equations for this system read
\begin{eqnarray}
H^2  &=& \frac{\kappa^2}{3}\,(\rho_{\rm dm}+\rho_f),
\label{Friedmann-1}\\[2mm]
2\dot{H}+3H^2 &=& -{\kappa}^2\,(p_c+p_f),
\label{Friedmann-2}
\end{eqnarray}
where $\kappa^2 = 8\pi G$, $H\equiv \dot{a}/a$ is the Hubble rate,
$\rho_{\rm dm}$ and $\rho_f$ denote the energy densities of DM and of the
second fluid, respectively, and $p_f$ is the pressure of the second fluid.
The quantity $p_c$ represents the creation pressure associated with
gravitationally induced particle production \cite{Steigman:2008bc,
Lima:2009ic, Basilakos:2010yp, deHaro:2015hdp} and is given by
\begin{equation}
p_c = -\frac{\Psi}{3H}\,\rho_{\rm dm},
\label{pcdefin}
\end{equation}
where $\Psi>0$ corresponds to matter creation and $\Psi<0$ describes particle
annihilation. The creation pressure emerges naturally from the
non-equilibrium thermodynamics of open systems, where an adiabatic
particle-production process, formulated within second-order relativistic
thermodynamics, leads to such an effective negative pressure consistent with
entropy conservation per particle.

Since the two fluids interact mutually and  exchange energy, the total 
conservation equation, namely
\begin{eqnarray}
\dot{\rho}_{\rm dm} + \dot{\rho}_f +
3H(\rho_{\rm dm}+\rho_f+p_c+p_f)=0,
\end{eqnarray}
can be decomposed as
\begin{eqnarray}\label{coupled}
\left\{
\begin{array}{l}
\dot{\rho}_{\rm dm} + 3H(\rho_{\rm dm}+p_c) = -Q, \\[2mm]
\dot{\rho}_f + 3H(\rho_f + p_f) = +Q,
\end{array}
\right.
\end{eqnarray}
where $Q$ denotes the interaction function describing the energy flow between
the fluids. Using (\ref{pcdefin}), the above equations gives
\begin{eqnarray}\label{mod-coupled}
\left\{
\begin{array}{l}
\dot{\rho}_{\rm dm} +
3H\Bigl(1 - \frac{\Psi}{3H}\Bigr)\rho_{\rm dm} = -Q, \\[2mm]
\dot{\rho}_f + 3H(1+w)\rho_f  = +Q,
\end{array}
\right.
\end{eqnarray}
where $w\equiv p_f/\rho_f$ is the barotropic equation-of-state (EoS) parameter
of the second fluid.

It is useful to rewrite \eqref{mod-coupled} in the form
\begin{eqnarray}\label{final-coupled}
\left\{
\begin{array}{l}
\dot{\rho}_{\rm dm} + 3H\left(1+w^{\rm eff}_{\rm dm}\right)\rho_{\rm dm}=0,
\\[2mm]
\dot{\rho}_f + 3H\left(1+w^{\rm eff}_{f}\right)\rho_f=0,
\end{array}
\right.
\end{eqnarray}
where the effective EoS parameters are
\begin{eqnarray}
w^{\rm eff}_{\rm dm}
= \frac{Q}{3H\rho_{\rm dm}} - \frac{\Psi}{3H},
\qquad
w^{\rm eff}_{f}
= w - \frac{Q}{3H\rho_f}.
\end{eqnarray}
Thus, both $w^{\rm eff}_{\rm dm}$ and $w^{\rm eff}_{f}$ may assume negative
values depending on the signs and magnitudes of the interaction rate $Q$ and
the matter-creation rate $\Psi$. 

At this point we would like to 
remark that there has been considerable interest on the possibility of   
non-cold 
DM~\cite{Muller:2004yb,Calabrese:2009zza,Kumar:2012gr,Armendariz-Picon:2013jej, 
Gariazzo:2017pzb,
Kopp:2018zxp,
Yao:2023ybs,Khurshudyan:2024gpn,Yang:2025ume,
Kumar:2025etf,Abedin:2025dis,Wang:2025zri,Chen:2025wwn,
Li:2025eqh,Yao:2025twv,Li:2025dwz,
Braglia:2025gdo,Yao:2025kuz}, and such a possibility could arise from the 
microscopic nature of DM though particle production or interaction between the 
fluids or both. In what follows we consider the simplest
matter-creation prescription, namely a constant rate $\Psi=\Psi_0$, and we
explore how the combined effect of matter creation and interaction modifies
the dynamical structure of the cosmology. This constant-rate assumption 
isolates the pure dynamical effect of interactions.
 
To this end, we investigate a representative set of well-known interaction
functions. We first consider the class of interactions proportional to the DM
energy density, namely
\begin{eqnarray}
A)~ Q = \mu\,H\rho_{\rm dm}, \qquad
B)~ Q = \Gamma\,\rho_{\rm dm},
\end{eqnarray}
where $\mu$ is dimensionless while $\Gamma$ has dimensions of $H$. These
choices are natural starting points, since the DM sector is the one endowed
with matter creation.

Next, motivated by the possibility that the energy flow may also depend on the
non-matter-creating fluid, we consider
\begin{eqnarray}
C)~ Q = \xi\,H\rho_f, \qquad
D)~ Q = \widetilde{\Gamma}\,\rho_f,
\end{eqnarray}
where $\xi$ is dimensionless and $\widetilde{\Gamma}$ has dimensions of $H$.

Finally, a symmetric extension of the above possibilities includes interaction
functions depending on both fluids:
\begin{eqnarray}
E)~ Q = \delta\,H\frac{\rho_{\rm dm}\rho_f}{\rho_{\rm dm}+\rho_f}, \qquad
F)~ Q = \overline{\Gamma}\,\frac{\rho_{\rm dm}\rho_f}
{\rho_{\rm dm}+\rho_f},
\end{eqnarray}
with $\delta$ dimensionless and $\overline{\Gamma}$ having the dimensions of
$H$.

In the above classification, Models A, C, and E involve a \emph{global}
interaction rate proportional to the instantaneous expansion rate $H$, whereas
Models B, D, and F correspond to \emph{local} interaction rates proportional to
a fixed scale with dimensions of $H$. These choices are dimensionally
consistent and lead to distinct dynamical behaviors in the resulting
autonomous systems.

In the next section we perform the phase-space analysis of each interacting
scenario, assuming constant matter creation in the DM sector, and show that
the combined presence of $Q$ and $\Psi_0$ enriches the dynamical structure
compared to standard interacting models without matter creation.

\section{Dynamical analysis}
\label{sec-dyn-analysis}

In order to perform the phase-space analysis, the gravitational equations are
rewritten as an autonomous dynamical system \cite{Bahamonde:2017ize}. To this 
end, we introduce the
dimensionless variables
\begin{equation}
x = \frac{\kappa^2 \rho_{\rm dm}}{3H^2}, 
\qquad
z = \frac{H_0}{H_0 + H},
\label{dimensionless-variables}
\end{equation}
where $H_0$ is the present value of the Hubble rate. Any positive constant
could be used in place of $H_0$, since its role is simply to compactify the 
phase
space along the $z$-direction. From Eq.~\eqref{Friedmann-1}, one obtains the
constraint
\[
x + \Omega_f = 1,
\qquad
\Omega_f \equiv \frac{\kappa^2 \rho_f}{3H^2},
\]
so that the physical phase space is the compact domain
\[
\mathbb{D}=\{(x,z): 0\le x \le 1,\; 0\le z \le 1\},
\]
with $z=0$ corresponding to $H\to\infty$ and $z=1$ corresponding to
$H\to 0$. Moreover, the deceleration parameter becomes
\[
q = \frac{1}{2}\!\left[1 + 3w(1-x) -
\frac{\alpha x z}{1-z}\right],
\]
where 
\begin{equation}
\alpha \equiv \frac{\Psi_0}{H_0} > 0
\end{equation}
is a dimensionless parameter encoding the constant matter-creation rate.
The term proportional to $\alpha$ represents the effective negative pressure
associated with adiabatic particle production, and it contributes as a
friction/anti-friction term in the dynamical evolution.

Using the field equations, the autonomous system valid for all interaction
models becomes
\begin{eqnarray}
x' &=& -\frac{\kappa^2 Q}{3H^3}
+ x(1-x)\left[3w + \frac{\alpha z}{1-z}\right],
\label{main-aut-1x}
\\[2mm]
z' &=& \frac{3}{2}z\left[(1+w(1-x))(1-z)
- \frac{\alpha x z}{3}\right],
\label{main-aut-1z}
\end{eqnarray}
where the prime denotes differentiation with respect to
$N=\ln(a/a_0)$, with $a_0$ the present scale factor.
For the interaction models $A$-$F$, the system is singular at $z=1$; this is
removed by the standard regularization procedure using the non-negative factor
$(1-z)$, following \cite{Bahamonde:2017ize,Halder:2024uao,
Halder:2024aan,Halder:2025eze}.

As we can see, the equation-of-state parameter $w$ of the second fluid enters 
the system
explicitly. Since matter creation alone can generate cosmic acceleration
without invoking an additional dark-energy component, our primary interest is
in non-negative values $w\ge 0$, while one can also consider $w<0$ 
in order to assess how a hypothetical negative-pressure fluid
would affect the dynamics of the interacting matter-creation scenarios. 
Although in the present article we have not considered such possibility, 
however, in a forthcoming article we shall present the results of such 
scenarios. Theoretically, the coexistence of matter creation and a dark energy 
component cannot be ruled out.  

In the following subsections we analyze the dynamical structure of each class
of interaction models.

 \subsection{Interactions proportional to $\rho_{dm}$  }
 
 \subsubsection{Model A: $Q = \mu H \rho_{\rm dm}$}

For the global interaction rate $Q=\mu H\rho_{\rm dm}$, the autonomous system
\eqref{main-aut-1x}-\eqref{main-aut-1z} becomes, after regularization,
\begin{eqnarray}
x' &=& -\mu x(1-z)
+ x(1-x)\left[3w(1-z)+\alpha z\right],
\label{rg-aut-1x}
\\[2mm]
z' &=& \frac{3}{2}\,z(1-z)
\left[(1+w(1-x))(1-z)-\frac{\alpha xz}{3}\right].
\label{rg-aut-1z}
\end{eqnarray}
  The surfaces $x=0$, $z=0$ and $z=1$ are invariant manifolds.  
On the boundary $x=1$, one finds $x'=-\mu(1-z)$, implying that for
$\mu>0$ the variable $x$ decreases near $x=1$.  
Therefore, the physical phase space $\mathbb{D}$ is positively invariant for  
\[
\alpha>0,\qquad \mu>0,
\]
which defines the admissible parameter region for this model.

The critical points, their properties, and the corresponding cosmological
quantities are summarized in Table~\ref{tab:model1-positive-w} (see \ref{sec-appendix}).  
Below we briefly describe their physical nature and stability.

\begin{figure}[!]
    \centering
    \includegraphics[width=0.38\textwidth]{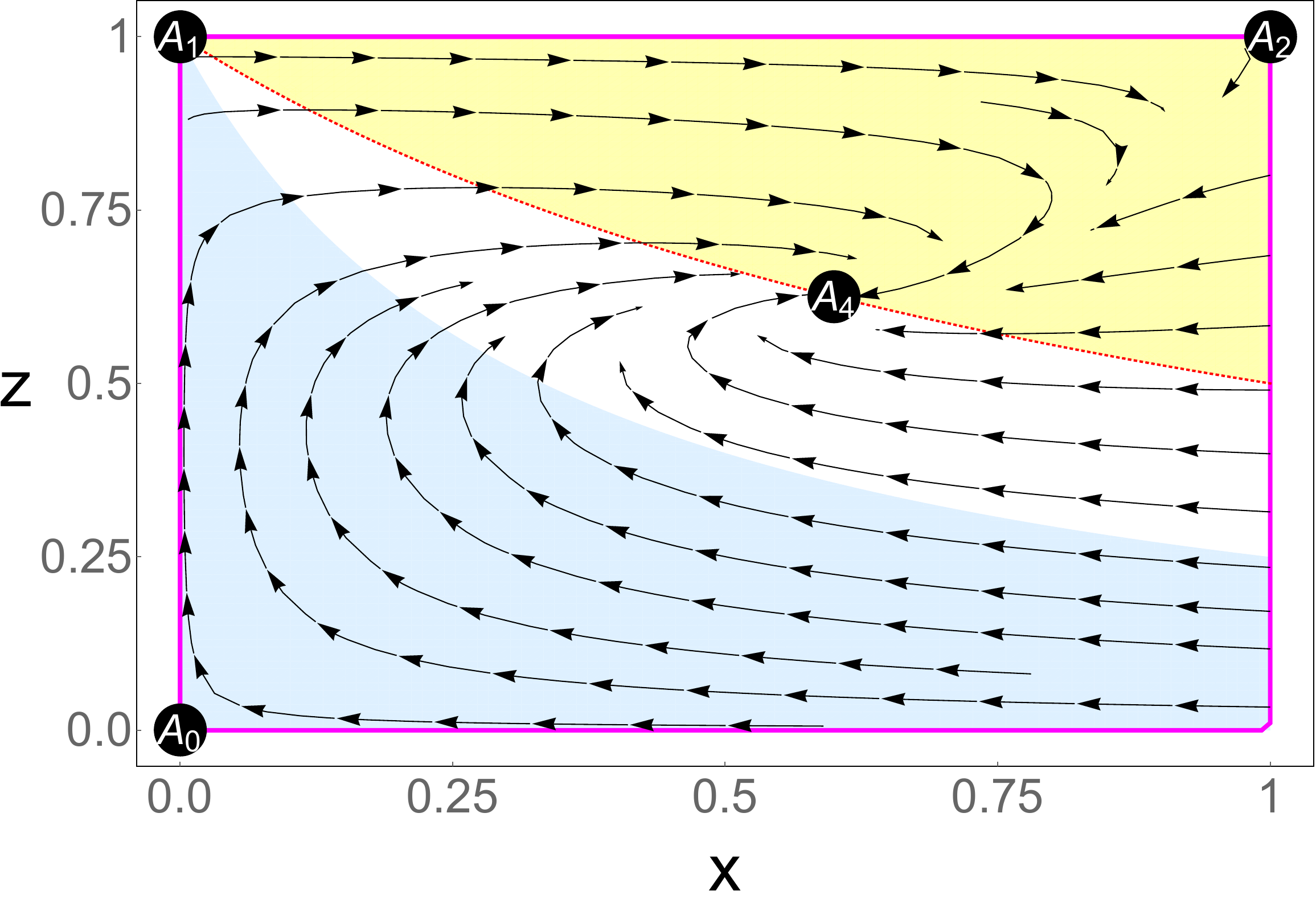}\\
    \includegraphics[width=0.38\textwidth]{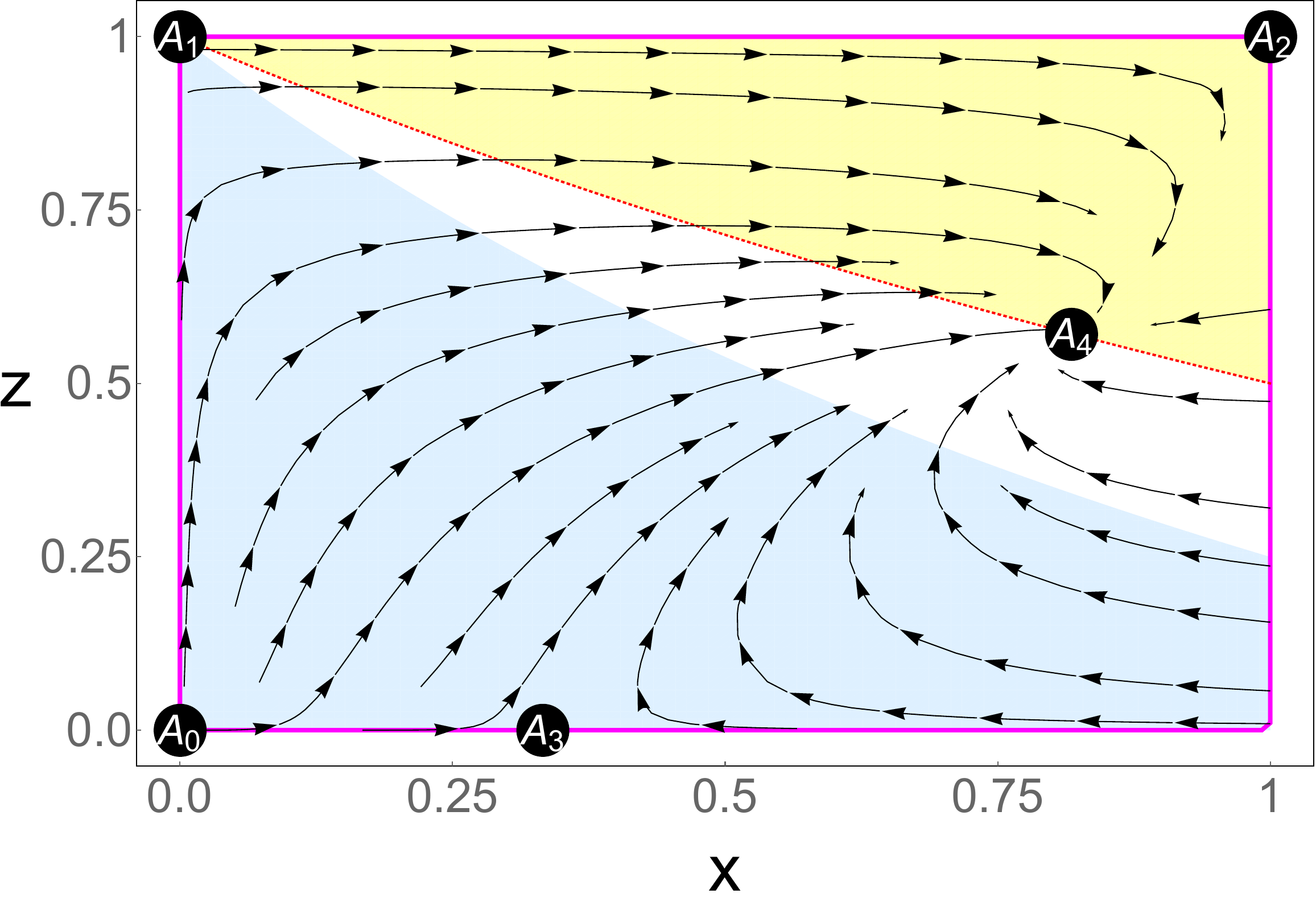}
    \caption{Phase space structure of the dynamical system 
(\ref{rg-aut-1x})-(\ref{rg-aut-1z}) with interaction rate $Q=\mu H \rho_{dm}$. 
{\bf Upper panel:} For $\alpha=3$, $w=0$, and $\mu=2$, the only attractor is 
$A_4$, while all other critical points are saddles. Similar behavior appears 
for 
$\alpha>0$, $\mu>0$, $w>0$ with $\mu>3w$. {\bf Lower panel:} For $\alpha=3$, 
$w=0.5$, and $\mu=1$, the attractor remains $A_4$, whereas the other critical 
points are saddles, except $A_0$, which is unstable. A similar structure is 
obtained when $\alpha>0$, $w>0$, $\mu>0$, and $\mu<3w$. The shaded regions 
indicate the cosmic expansion phases: blue for deceleration $(q>0)$, white for 
acceleration $(-1<q<0)$, and yellow for super-acceleration $(q<-1)$. The red 
dotted curve marks the boundary $q=-1$.  }
    \label{fig1}
\end{figure}

\texttt{(i)} \textbf{A$_0$}:  
A second-fluid-dominated point ($\Omega_f=1$) with saddle character for
$\mu\ge 3w$.  
Since $w\ge0$, this point corresponds to a decelerating early-time state. \texttt{(ii)} \textbf{A$_1$}:  
A second-fluid-dominated critical point with non-hyperbolic saddle nature.
The deceleration parameter is undefined here, making the phase ambiguous from
a physical viewpoint. \texttt{(iii)} \textbf{A$_2$}:  
A DM-dominated point ($\Omega_{\rm dm}=1$) with saddle stability.  
It corresponds to an accelerating state, but cannot act as a late-time
attractor. \texttt{(iv)} \textbf{A$_3$}:  
A coexistence point where both fluids are present.  
It is a saddle and always leads to decelerating expansion. \texttt{(v)} \textbf{A$_4$}:  
The most relevant point in this model.  
Both fluids coexist, the point is stable, and it yields an accelerating
late-time expansion (see Fig.~\ref{fig1}).  
It represents an accelerating scaling attractor.

\begin{figure}
    \includegraphics[width=0.43\textwidth]{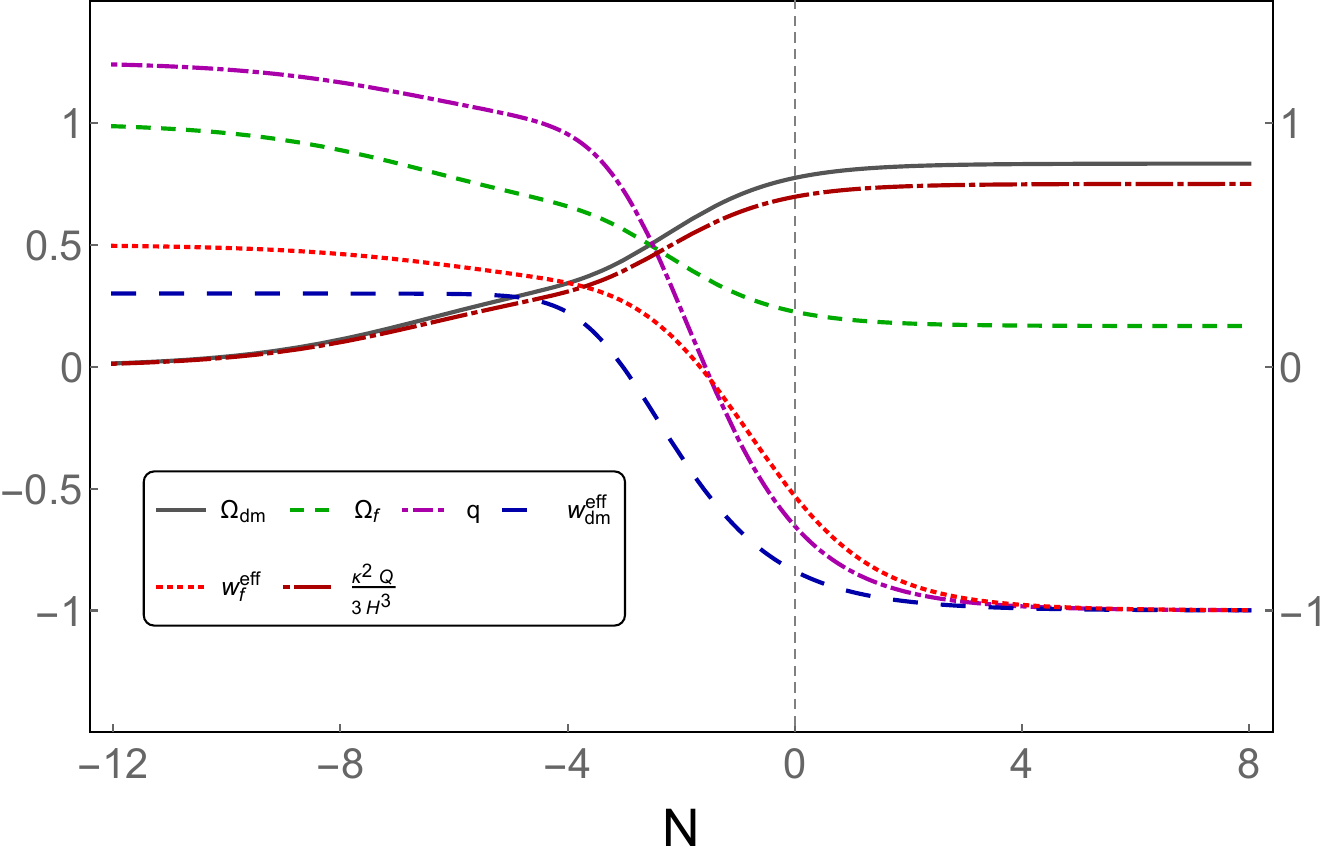}
    \caption{Evolution of the DM density parameter $(\Omega_{dm})$, second 
fluid 
density parameter $(\Omega_f)$, deceleration parameter $(q)$, effective EoS of 
DM $(w^{\rm eff}_{dm})$, effective EoS of the second fluid $(w^{\rm eff}_f)$, 
and 
coupling strength $(\frac{\kappa^2 Q}{3H^3})$ with respect to the e-folding 
number $N$ for the dynamical system (\ref{rg-aut-1x})-(\ref{rg-aut-1z}) 
associated with the interaction rate $Q=\mu H \rho_{\rm dm}$. The plot is 
obtained 
for $\alpha=1.5$, $w=0.5$, $\mu=0.9$ with initial conditions $x(-9)=0.07$, 
$z(-9)=5.0\times 10^{-6}$.   }
    \label{fig2}
\end{figure}

Fig. \ref{fig2} displays the evolution of $\Omega_{\rm dm}$,
$\Omega_f$, $q$, $w^{\rm eff}_{\rm dm}$, $w^{\rm eff}_{f}$ and
$\kappa^2 Q/(3H^3)$.  
At late times, an energy flow from DM to the second fluid drives both density
parameters toward constant non-zero asymptotic values, resulting in accelerated
expansion.  
The interaction and matter creation together force the effective equations of
state to negative values, consistent with the emergence of the attractor A$_4$.

Finally, if $\mu<0$ (energy transfer from the second fluid to DM), the physical
phase space is no longer positively invariant and the accelerating scaling
attractor disappears.  
Thus, the existence of A$_4$ requires a net energy flow from DM to the second
fluid.

A final remark is in order: 
even if we include baryons and radiation in the Hubble equation where both of them evolve independently, 
the qualitative nature of this interacting matter creation scenario is not altered. Hence, the existence or stability of the accelerating scaling attractors does not get affected in this new extended scenario. However, the extended dynamical system admits a radiation-dominated phase at early times, followed by a decelerating matter-dominated era prior to the onset of late-time acceleration, thereby preserving the standard cosmological sequence.

\subsubsection{Model B: $Q = \Gamma \rho_{\rm dm}$}

For the local interaction rate $Q=\Gamma\rho_{\rm dm}$, the regularized
autonomous system \eqref{main-aut-1x}-\eqref{main-aut-1z} becomes
\begin{eqnarray}
x' &=& -\beta x z
+ x(1-x)\left[3w(1-z)+\alpha z\right],
\label{rg-aut-2x}
\\[2mm]
z' &=& \frac{3}{2}\,z(1-z)\left[(1+w(1-x))(1-z)
- \frac{\alpha x z}{3}\right],
\label{rg-aut-2z}
\end{eqnarray}
where $\beta \equiv \Gamma/H_0$ is a dimensionless coupling constant.
As in Model A, the hypersurfaces $x=0$, $z=0$, and $z=1$ are invariant
manifolds. On the boundary $x=1$, one obtains $x'=-\beta z$, implying that
$x$ decreases for $\beta>0$.  
Thus, the physical domain $\mathbb{D}$ remains positively invariant for
\[
\alpha>0, \qquad \beta>0,
\]
which defines the admissible parameter space of this scenario.

The critical points and their cosmological properties are listed in
Table~\ref{tab:model1-positive-w} (see \ref{sec-appendix}).   
Their physical interpretation is summarized below.

\begin{figure}[!]
    \centering
    \includegraphics[width=0.38\textwidth]{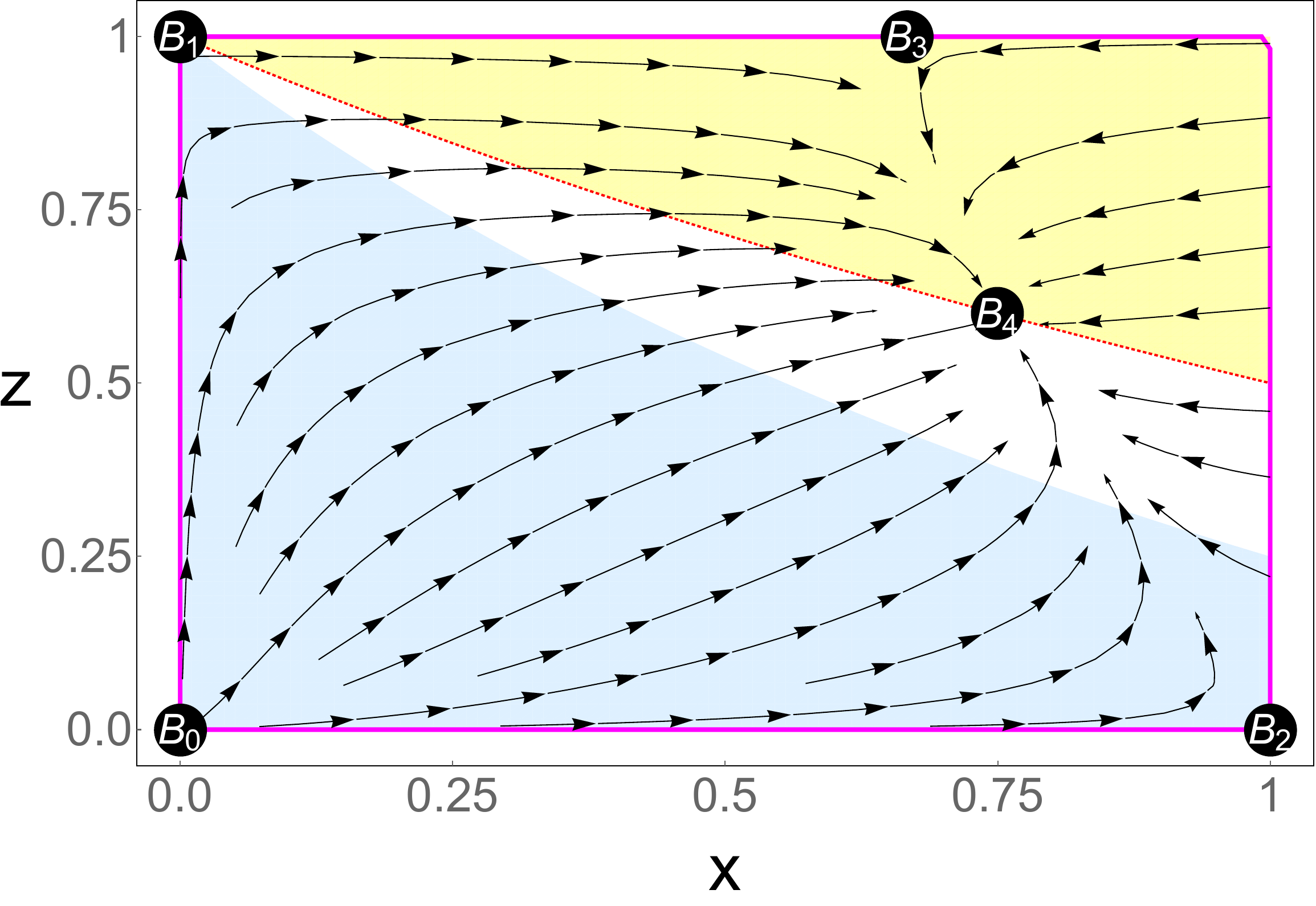}\\
    \includegraphics[width=0.38\textwidth]{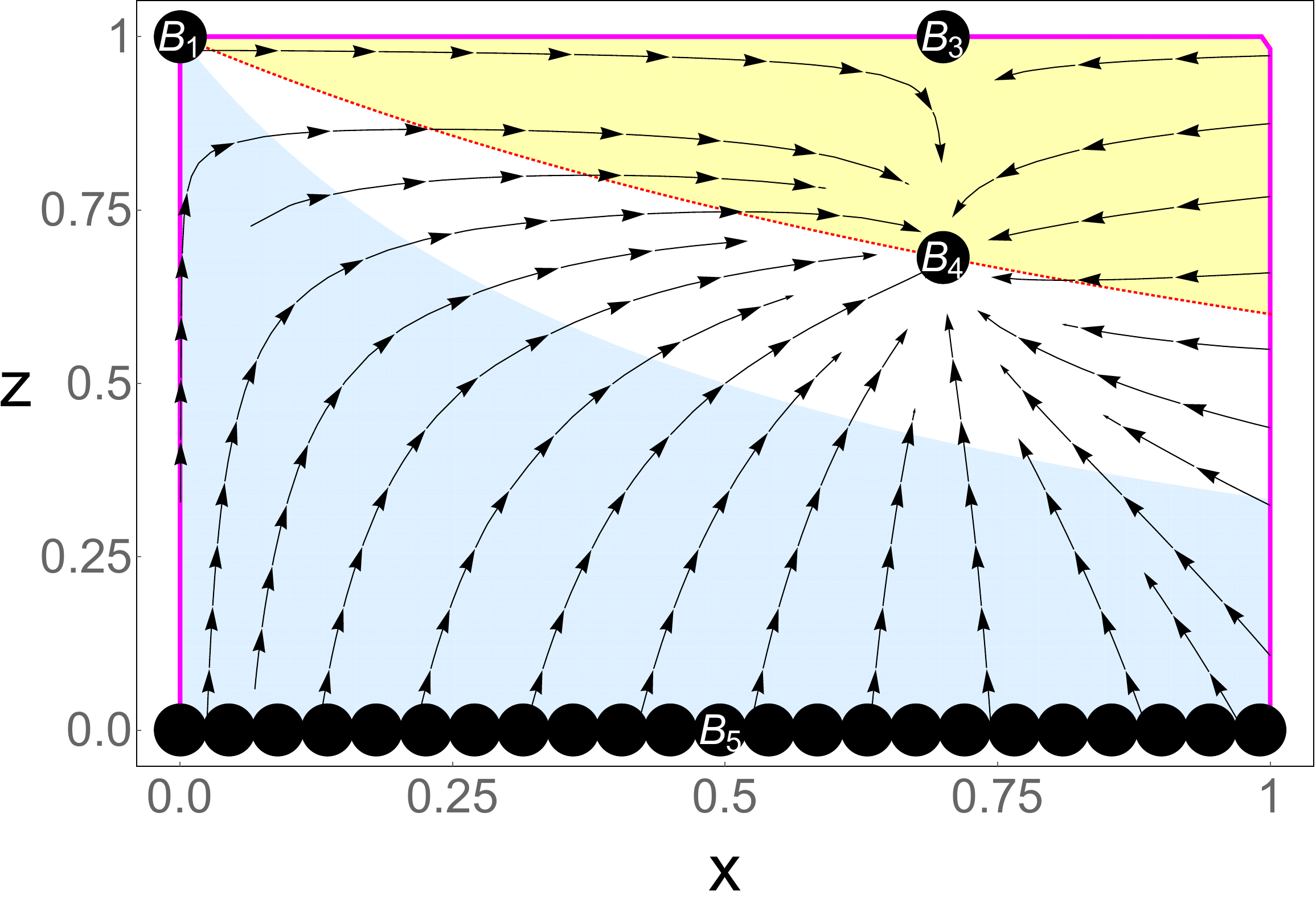}
    \caption{Phase space diagrams of the dynamical system 
(\ref{rg-aut-2x})-(\ref{rg-aut-2z}) with the interaction rate $Q=\Gamma 
\rho_{\rm dm}$. {\bf Upper panel:} For $\alpha=3$, $w=0.5$, and $\beta=1$, the 
only 
attractor is $B_4$, while all other critical points are saddles, except $B_0$, 
which is unstable. Similar behavior arises for $\alpha>0$, $\beta>0$, $w>0$ 
with 
$\alpha>\beta$. {\bf Lower panel:} For $\alpha=2$, $w=0$ and $\beta=0.6$, the 
attractor remains $B_4$; the points $B_1$ and $B_3$ are saddles, while the 
critical line $B_5$ is unstable. Comparable structures occur for $\alpha>0$, 
$w=0$, $\beta>0$ with $\alpha>\beta$. Blue, white, and yellow regions 
correspond 
to deceleration $(q>0)$, acceleration $(-1<q<0)$, and super-acceleration 
$(q<-1)$, respectively. The red dotted curve marks the transition curve $q=-1$. 
 }
    \label{fig3}
\end{figure}

\begin{figure}
    \includegraphics[width=0.43\textwidth]{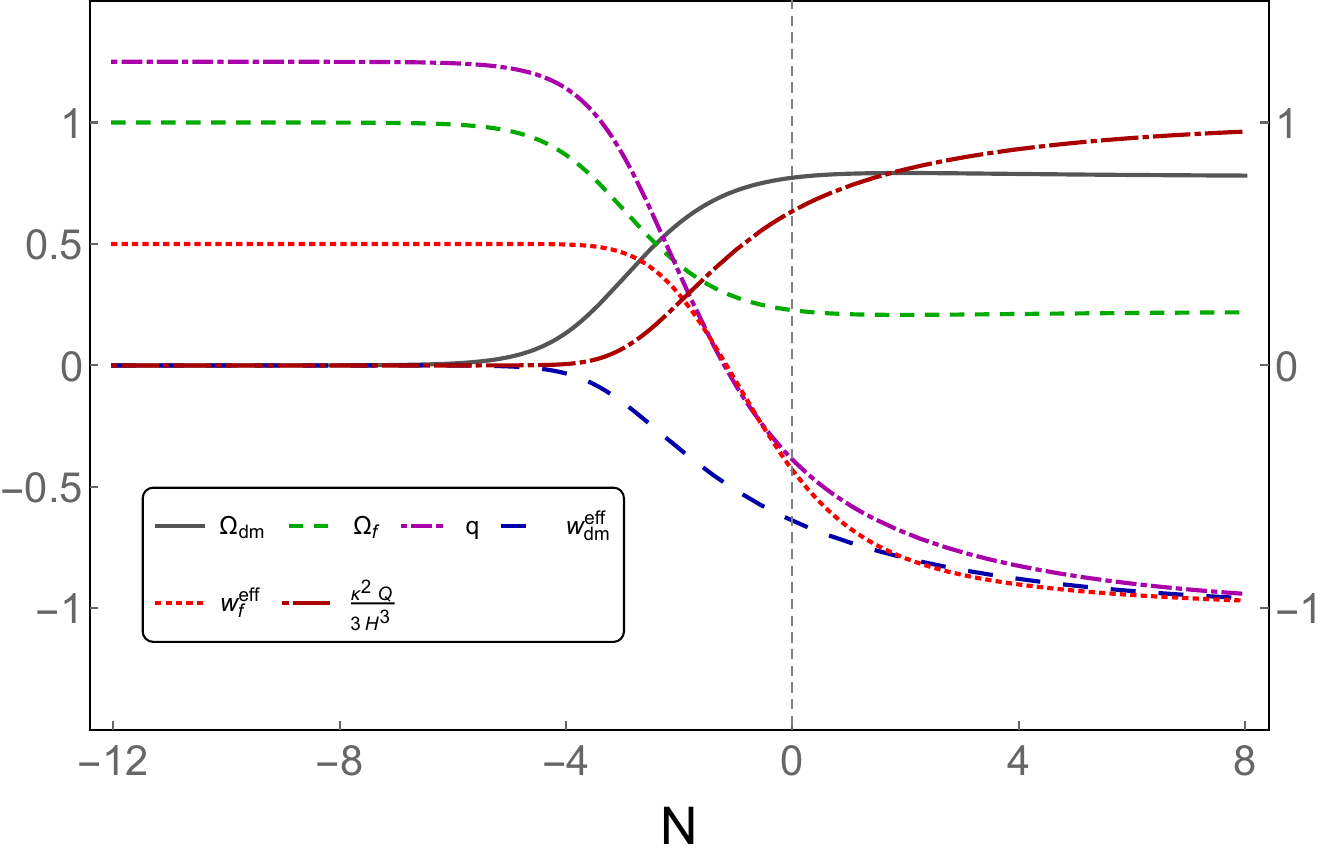}
    \caption{Dynamics of the system (\ref{rg-aut-2x})-(\ref{rg-aut-2z}) with 
interaction rate $Q=\Gamma \rho_{\rm dm}$, showing the evolution of 
$\Omega_{\rm dm}$, 
$\Omega_f$, decelerating parameter $q$, the effective EoS parameters 
$w^{\rm eff}_{\rm dm}$ and $w^{\rm eff}_f$, and the coupling strength 
$\frac{\kappa^2 
Q}{3H^3}$ as a function of the e-folding number $N$. The trajectory is 
computed using $\alpha=1$, $w=0.5$, $\beta=0.3$, with initial conditions 
$x(-4.5)=0.07$, $z(-4.5)=5.0\times 10^{-2}$. }
    \label{fig4}
\end{figure}

\texttt{(i)} \textbf{B$_0$}:  
A second-fluid-dominated ($\Omega_f=1$) point describing a decelerating and
unstable early-time state. \texttt{(ii)} \textbf{B$_1$}:  
Also second-fluid-dominated.  
Its stability depends on the relative values of $\alpha$ and $\beta$: it is a
saddle for $\alpha>\beta$ and stable for $\alpha\le\beta$.  
The deceleration parameter is undefined at this point, leaving its physical
interpretation ambiguous. \texttt{(iii)} \textbf{B$_2$}:  
A DM-dominated point ($\Omega_{\rm dm}=1$) describing a decelerating era.
It is a saddle for $w>0$, while for $w=0$ it becomes unstable. \texttt{(iv)} \textbf{B$_3$}:  
A coexistence point of the two fluids with $q<0$, implying accelerated
expansion.  
It is stable for $\alpha=\beta$ and a saddle for $\alpha>\beta$. \texttt{(v)} \textbf{B$_4$}:  
A physically relevant coexistence point where both fluids remain present.
It is stable and corresponds to an accelerating late-time state (see
Fig.~\ref{fig3}).  
This point generically plays the role of an accelerating scaling attractor. \texttt{(vi)} \textbf{B$_5$}:  
A non-hyperbolic, unstable, decelerating critical point.  
Its nature depends on the value of $x_c$:  
for $x_c=0$ the universe is second-fluid-dominated,  
for $x_c=1$ it is DM-dominated,  
and for $0<x_c<1$ both fluids coexist.

Fig. \ref{fig4} illustrates the evolution of $\Omega_{\rm dm}$,
$\Omega_f$, $q$, $w^{\rm eff}_{\rm dm}$, $w^{\rm eff}_{f}$, and
$\kappa^2 Q/(3H^3)$.  
At late times, an energy flow from DM to the second fluid drives the system
toward accelerated expansion, with both density parameters asymptoting to
non-zero constants.  
The interaction strength and matter-creation rate jointly push the effective
equations of state into the negative regime, in agreement with the existence of
the stable accelerating attractor B$_4$.

As in the previous model, a reversal of the energy flow ($\beta<0$) destroys the
positive invariance of the phase space and eliminates the accelerating scaling
attractor.  
Thus, accelerated scaling solutions require a net transfer of energy from DM to
the second fluid.

In analogy with Model A, the inclusion of radiation and baryons preserves the standard cosmological sequence, featuring a radiation-dominated epoch followed by a decelerating DM-dominated phase prior to late-time acceleration, while the existence and stability of the accelerating scaling attractors remain unaffected.

 \subsection{Interactions proportional to $\rho_{f}$  (Models C and D)}

 \subsubsection{Model C: $Q = \xi H \rho_f$}

For the global interaction rate $Q=\xi H\rho_f$, the regularized autonomous
system \eqref{main-aut-1x}-\eqref{main-aut-1z} becomes
\begin{eqnarray}
x' &=& (1-x)\left[-\xi(1-z) + 3w x (1-z) + \alpha x z \right],
\label{rg-aut-3x}
\\[2mm]
z' &=& \frac{3}{2}\,z(1-z)\left[(1+w(1-x))(1-z) - \frac{\alpha x z}{3}\right].
\label{rg-aut-3z}
\end{eqnarray}
From these equations one identifies $x=1$, $z=0$, and $z=1$ as invariant
manifolds.  
Along the boundary $x=0$ we have $x' = -\xi(1-z)$; thus, $x$ increases near
$x=0$ provided $\xi<0$.  
Consequently, the physical phase space $\mathbb{D}$ is positively invariant for
\[
\alpha>0, \qquad \xi<0,
\]
which therefore defines the admissible parameter region of this model.

The critical points and their cosmological properties are listed in
Table~\ref{tab:model2-positive-w} (see \ref{sec-appendix}).  
Their physical interpretation is summarised below.

\texttt{(i)} \textbf{C$_0$}:  
A second-fluid-dominated, non-hyperbolic saddle.  
The deceleration parameter is undefined at this point, preventing a definite
physical characterization.
\texttt{(ii)} \textbf{C$_1$}:  
A DM-dominated saddle corresponding to a decelerating state.
\texttt{(iii)} \textbf{C$_2$}: 
A DM-dominated saddle representing a super-accelerating phase
($q< -1$).  
Despite its accelerating nature, it cannot act as a late-time attractor. \texttt{(iv)} \textbf{C$_3$}:  
The only stable point in this model.  
It corresponds to accelerated expansion but remains entirely DM-dominated
($\Omega_{\rm dm}=1$), with no coexistence between the two fluids.

In contrast to Models A and B, no accelerating scaling solution emerges in this
scenario: even though energy exchange occurs between DM and the second fluid,
all accelerating solutions are either saddle points or fully DM-dominated
attractors.  
Thus, interactions proportional to $\rho_f$ do not support a stable scaling
accelerating state. 

Lastly, we also noticed that the inclusion of baryons and radiation does not alter the overall qualitative nature of this interacting matter creating model. The resulting dynamics naturally exhibits a radiation-dominated epoch at early times, followed by a decelerating matter-dominated phase preceding the onset of late-time acceleration, thus preserving the standard cosmological sequence. However, no accelerating scaling attractors are found; rather, the late-time evolution is characterized by an accelerating dark matter-dominated phase.

 \subsubsection{Model D: $Q = \widetilde{\Gamma}\,\rho_f$}

For the local interaction rate $Q=\widetilde{\Gamma}\rho_f$, the regularized
autonomous system \eqref{main-aut-1x}-\eqref{main-aut-1z} becomes
\begin{eqnarray}
x' &=& -\gamma (1-x) z
+ x(1-x)\left[3w(1-z)+\alpha z\right],
\label{rg-aut-4x}
\\[2mm]
z' &=& \frac{3}{2}\,z(1-z)\left[(1+w(1-x))(1-z)
- \frac{\alpha x z}{3}\right],
\label{rg-aut-4z}
\end{eqnarray}
where $\gamma \equiv \widetilde{\Gamma}/H_0$ is a dimensionless coupling
parameter.  
The surfaces $x=1$, $z=0$, and $z=1$ are invariant manifolds.  
Along the line $x=0$ we have $x'=-\gamma z$, so $x$ increases near $x=0$ when
$\gamma<0$.  
Thus, the physical phase space $\mathbb{D}$ is positively invariant provided
\[
\alpha>0, \qquad \gamma<0,
\]
which defines the allowed parameter region for this model.

The critical points and their cosmological properties are summarized in
Table~\ref{tab:model2-positive-w} (see \ref{sec-appendix}).  
Their physical interpretation is as follows.

\texttt{(i)} \textbf{D$_0$}:  
A second-fluid-dominated, decelerating state ($q>0$) with two positive
eigenvalues and hence unstable.
\texttt{(ii)} \textbf{D$_1$}:  
A DM-dominated saddle describing a decelerating universe.
\texttt{(iii)} \textbf{D$_2$}: 
A DM-dominated saddle associated with accelerated expansion.  
Although accelerating, this point cannot serve as a late-time attractor.
\texttt{(iv)} \textbf{D$_3$}:  
The unique stable critical point in the phase space.  
It corresponds to an accelerating state but remains fully DM-dominated
($\Omega_{\rm dm}=1$), with no contribution from the second fluid.
\texttt{(v)} \textbf{D$_4$}:  
A critical line in which both fluids are present (except at $x_c=0,1$).  
With one zero and one positive eigenvalue, it is non-hyperbolic and unstable,
corresponding to a decelerating evolution.

As in Model C, interactions proportional to $\rho_f$ do not generate an
accelerating scaling attractor: all accelerating solutions are either saddles or
completely DM-dominated late-time attractors.  
Thus, even allowing for energy flow in either direction, this class of
interactions cannot sustain a stable accelerating state with coexistence of both
fluids.

 Similarly to Model C, the inclusion of  radiation and baryons maintains the standard cosmological sequence. However, no accelerating scaling attractors are found, and the late-time evolution leads to an accelerating DM-dominated epoch.

   \subsection{Mixed interactions (Models E and F)}
   
    \subsubsection{Model E: $Q = \delta H \dfrac{\rho_{\rm dm}\rho_f}{\rho_{\rm 
dm}+\rho_f}$}

For the global interaction rate $Q=\delta H\dfrac{\rho_{\rm 
dm}\rho_f}{\rho_{\rm 
dm}+\rho_f}$,
the regularized autonomous system \eqref{main-aut-1x}-\eqref{main-aut-1z}
takes the form
\begin{eqnarray}
x' &=& x(1-x)\Bigl[-\delta(1-z) + 3w(1-z) + \alpha z\Bigr],
\label{rg-aut-5x}
\\[2mm]
z' &=& \frac{3}{2}\,z(1-z)
\left[(1+w(1-x))(1-z)-\frac{\alpha xz}{3}\right].
\label{rg-aut-5z}
\end{eqnarray}
The boundaries $x=0$, $x=1$, $z=0$, and $z=1$ are invariant manifolds, ensuring
that the compact phase space $\mathbb{D}$ is positively invariant for all
$\alpha>0$ and $\delta$ of either sign.  
The corresponding critical points and their cosmological properties are listed
in Table~\ref{tab:model3-positive-w} (see \ref{sec-appendix}).  
We summarize their physical interpretation below.

\begin{figure}[ht]
    \centering
    \includegraphics[width=0.42\textwidth]{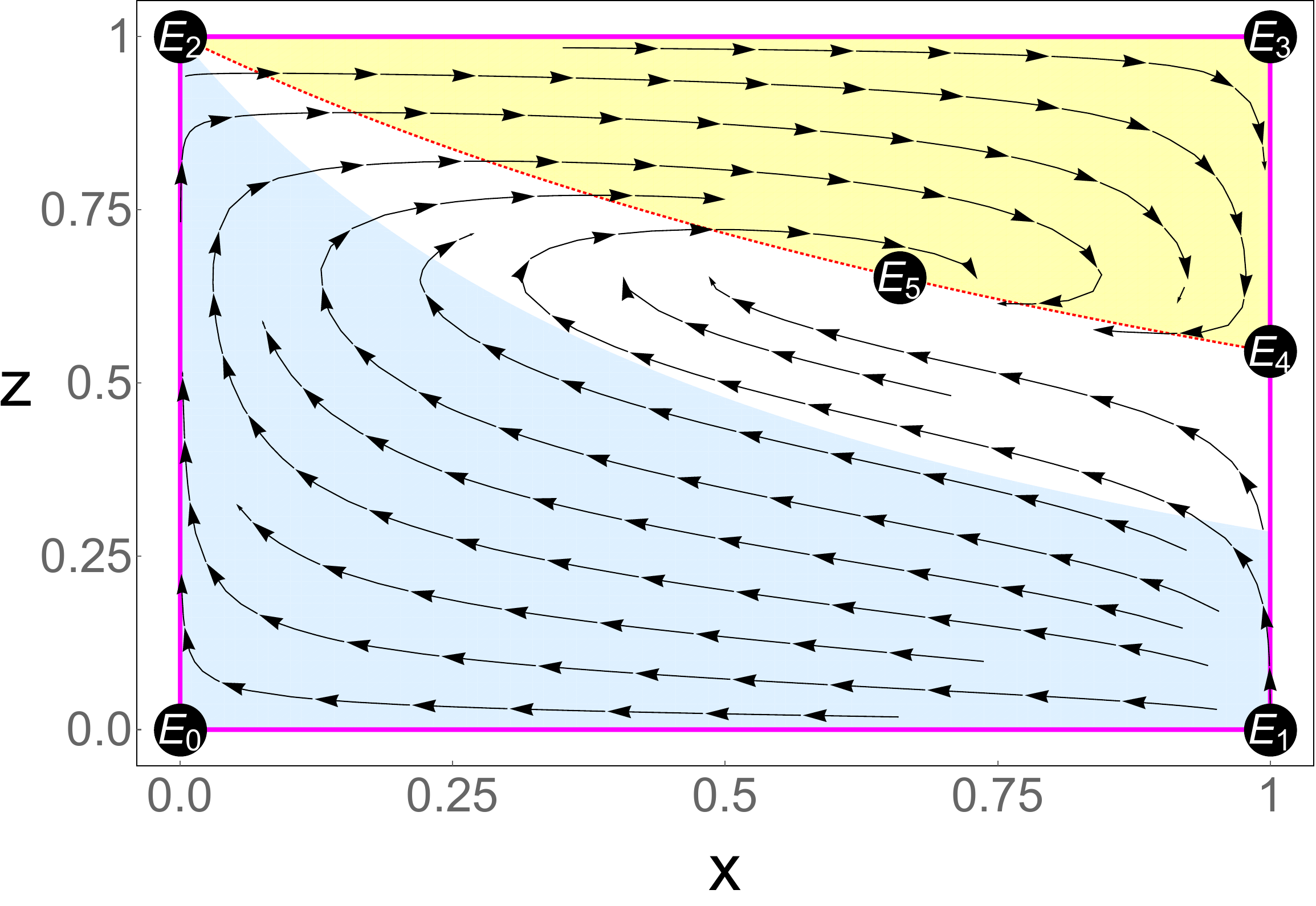}
    \caption{Phase portrait of the dynamical system 
(\ref{rg-aut-5x})-(\ref{rg-aut-5z}) for the interaction rate $Q=\delta H 
\rho_{dm}\rho_{f}(\rho_{dm}+\rho_{f})^{-1}$. For $\alpha=2.5$, $w=0.1$, and 
$\delta=5$, the critical point $E_5$ emerges as the only attractor, while all 
other critical points are saddles, except $E_1$, which is unstable. Similar 
phase portraits appear for $\alpha>0$, $\delta>0$, $w\geq0$ with 
$\delta>3(1+w)$. Blue, white, and yellow zones represent decelerating $(q>0)$, 
accelerating $(-1<q<0)$, and super-accelerating $(q<-1)$ regimes, respectively. 
The red dotted curve denotes the boundary $q=-1$.  }
    \label{fig5}
\end{figure}

\texttt{(i)} \textbf{E$_0$}:  
A second-fluid-dominated ($\Omega_f=1$) decelerating state.  
It is a saddle for $\delta>3w$ and unstable for $\delta<3w$. \texttt{(ii)} \textbf{E$_1$}:  
A DM-dominated ($\Omega_{\rm dm}=1$) decelerating critical point.  
It is unstable for $\delta>3w$ and a saddle for $\delta<3w$. \texttt{(iii)} \textbf{E$_2$}:  
A second-fluid-dominated, non-hyperbolic saddle with undefined deceleration
parameter, preventing any physical interpretation regarding acceleration. \texttt{(iv)} \textbf{E$_3$}:  
A DM-dominated saddle corresponding to an accelerating universe. \texttt{(v)} \textbf{E$_4$}:  
A DM-dominated accelerating point with $q=-1$.  
It is stable for $\delta\le 3(w+1)$ and a saddle for $\delta>3(w+1)$. \texttt{(vi)} \textbf{E$_5$}:  
The most relevant point in this model.  
It is stable, accelerating ($q=-1$), and features the coexistence of both
fluids ($0<x_c<1$).  
Thus, E$_5$ represents an accelerating scaling attractor (see
Fig.~\ref{fig5}). \texttt{(vii)} \textbf{E$_6$}:  
An unstable, decelerating critical point.  
Depending on $x_c$, it may be DM-dominated ($x_c=1$), second-fluid-dominated
($x_c=0$), or a coexistence state ($0<x_c<1$).

\begin{figure}
    \includegraphics[width=0.44\textwidth]{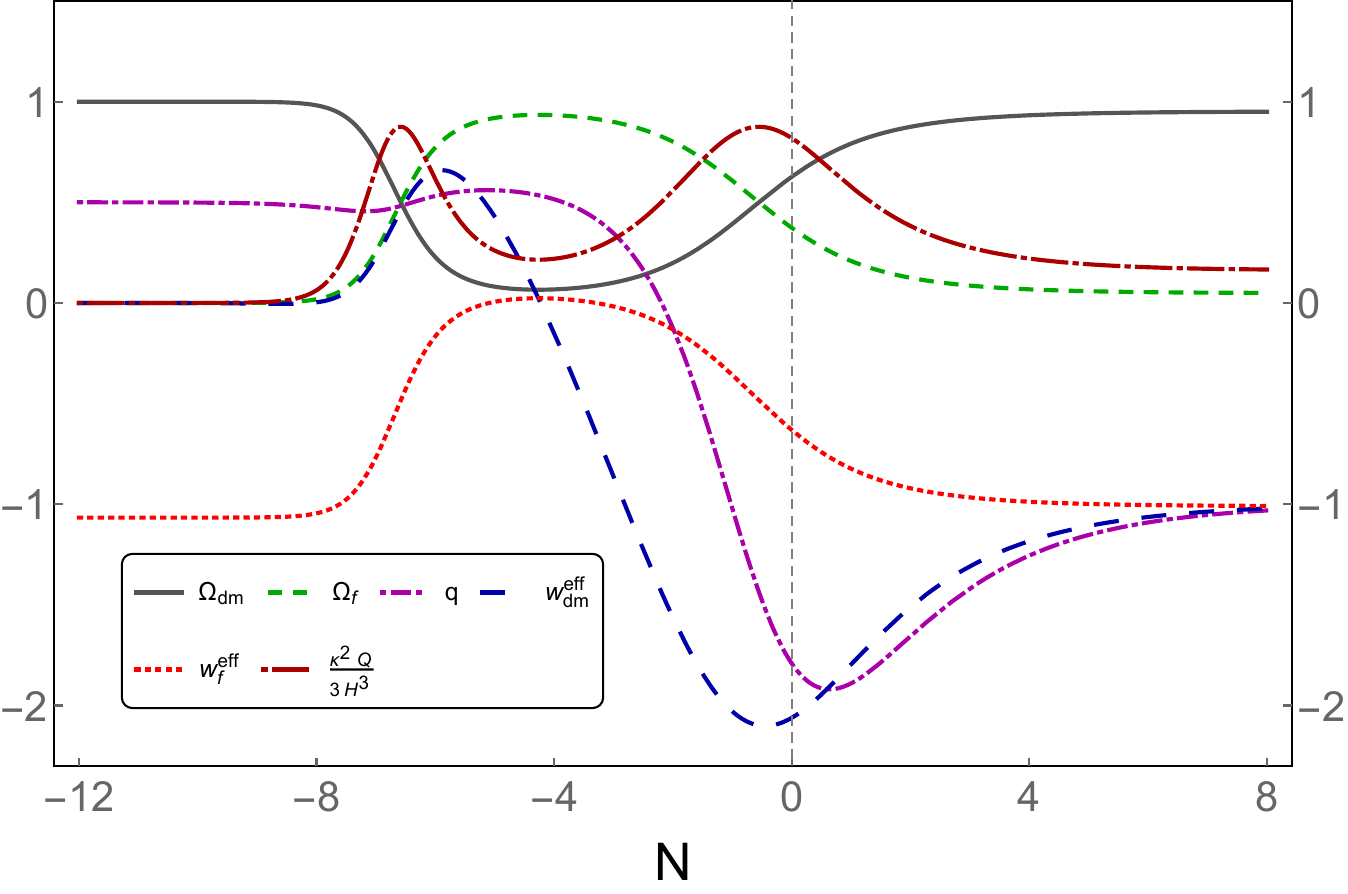}
    \caption{Evolution of $\Omega_{\rm dm}$, $\Omega_f$, $q$, $w^{\rm eff}_{\rm 
dm}$, 
$w^{\rm eff}_f$, and $\frac{\kappa^2 Q}{3H^3}$ versus e-folding number $N$, 
for the dynamical system (\ref{rg-aut-5x})-(\ref{rg-aut-5z}) with the 
interaction rate $ Q = \delta H \rho_{\rm dm} \rho_{f}(\rho_{\rm dm} + 
\rho_{f})^{-1} $. Parameters: $\alpha=2$, $w=0.1$, $\delta=3.5$. Initial 
conditions: $x(-8.2)=0.99$, $z(-8.2)=2.0\times 10^{-2}$.}
    \label{fig6}
\end{figure}

Fig. \ref{fig6} shows the late-time behaviour of the system.  
The universe undergoes accelerated expansion while the energy density is shared
between the two fluids, driven by a net energy flow from DM to the second fluid.
The interaction and matter-creation contributions jointly push the effective
equations of state $w^{\rm eff}_{\rm dm}$ and $w^{\rm eff}_{f}$ into the
negative regime, consistent with the stability of E$_5$.

Finally, we note that the accelerating scaling attractor occurs only when
$\delta>0$, corresponding to energy transfer from DM to the second fluid.
When $\delta<0$, the late-time attractor disappears and no stable accelerating
scaling solution exists.

As in Models A and B, including radiation and baryons preserves the standard cosmological sequence and the qualitative nature of the model does not change. Accelerating scaling attractors obtained in this new scenario remain qualitatively unchanged in both existence and stability.

\subsubsection{Model F: $Q = \overline{\Gamma}\,\dfrac{\rho_{\rm 
dm}\rho_f}{\rho_{\rm dm}+\rho_f}$}

For the local interaction rate 
\(
Q=\overline{\Gamma}\,\dfrac{\rho_{\rm dm}\rho_f}{\rho_{\rm dm}+\rho_f},
\)
the regularized autonomous system \eqref{main-aut-1x}-\eqref{main-aut-1z}
reduces to
\begin{eqnarray}
x' &=& x(1-x)\Bigl[-\nu z + 3w(1-z) + \alpha z\Bigr],
\label{rg-aut-6x}
\\[2mm]
z' &=& \frac{3}{2}\,z(1-z)
\left[(1+w(1-x))(1-z) - \frac{\alpha xz}{3}\right],
\label{rg-aut-6z}
\end{eqnarray}
where $\nu \equiv \overline{\Gamma}/H_0$ is a dimensionless coupling parameter.
The hypersurfaces $x=0$, $x=1$, $z=0$, and $z=1$ are invariant manifolds, so the
compact domain $\mathbb{D}$ remains positively invariant for all admissible
parameter values.  
The corresponding critical points and their stability properties are listed in
Table~\ref{tab:model3-positive-w} (see \ref{sec-appendix}).  
Their physical interpretation is as follows.

\texttt{(i)} \textbf{F$_0$}:  
A second-fluid-dominated ($\Omega_f=1$) decelerating state.  
Both eigenvalues are positive, so the point is unstable. \texttt{(ii)} \textbf{F$_1$}:  
A DM-dominated ($\Omega_{\rm dm}=1$) saddle describing a decelerating phase. \texttt{(iii)} \textbf{F$_2$}:  
A second-fluid-dominated, non-hyperbolic critical point.  
It is  saddle when $\nu<\alpha$ and stable when $\nu>\alpha$.  
The deceleration parameter is undefined, preventing further interpretation. \texttt{(iv)} \textbf{F$_3$}:  
A DM-dominated accelerating solution.  
It is stable for $\nu<\alpha$ and unstable for $\nu>\alpha$. \texttt{(v)} \textbf{F$_4$}:  
Another DM-dominated accelerating state.  
It is stable for $\nu\le\alpha(1+w)$ and a saddle for $\nu>\alpha(1+w)$. \texttt{(vi)} \textbf{F$_5$}:  
An accelerating coexistence point ($0<x_c<1$), but of saddle type. \texttt{(vii)} \textbf{F$_6$}:  
An accelerating non-hyperbolic point that may be DM-dominated ($x_c=1$),
second-fluid-dominated ($x_c=0$), or a coexistence state ($0<x_c<1$).  
It is always unstable. \texttt{(viii)} \textbf{F$_7$}:  
A fully unstable decelerating critical point.  
Depending on $x_c$, it may correspond to DM dominance, second-fluid dominance,
or coexistence of both fluids.


Overall, Model F shares the qualitative features of Model E, but with the
crucial distinction that none of its accelerating coexistence points are stable.
Energy exchange proportional to the mixed density term does not yield a stable
accelerating scaling attractor: all stable late-time accelerating solutions are
fully DM-dominated.

As in Models C and D, including radiation and baryons retains the standard cosmological sequence. No accelerating scaling attractors exist, and the Universe approaches an accelerating DM-dominated epoch.

\section{Comparison of Interaction Models}
\label{sec-comparison}

In the previous section we performed a detailed analysis of 
accelerating scaling attractors. Hence, it is useful to summarize the 
qualitative differences among the interaction classes considered above. 
Despite their distinct functional forms, most interaction prescriptions lead to 
qualitatively similar late-time dynamics, typically driving the system either 
toward dark-matter domination or toward fixed points where the effective 
equation of state fails to produce accelerated expansion. 
In particular, interactions proportional to the density of the secondary fluid 
or to combinations dominated by it do not admit accelerating scaling solutions, 
independently of parameter choices. 
On the other hand, interaction terms in which the energy transfer is 
effectively controlled by the dark-matter sector exhibit a richer dynamical 
structure and can, under suitable conditions, support late-time attractors with 
both acceleration and a constant ratio of energy densities. 
This classification motivates a more focused investigation of the precise 
conditions under which accelerating scaling attractors can emerge, which we 
address in the following.

\subsection{Conditions for accelerating scaling attractors}

\begin{figure}[ht]
\hspace{-1cm}
    \includegraphics[width=0.5\textwidth]{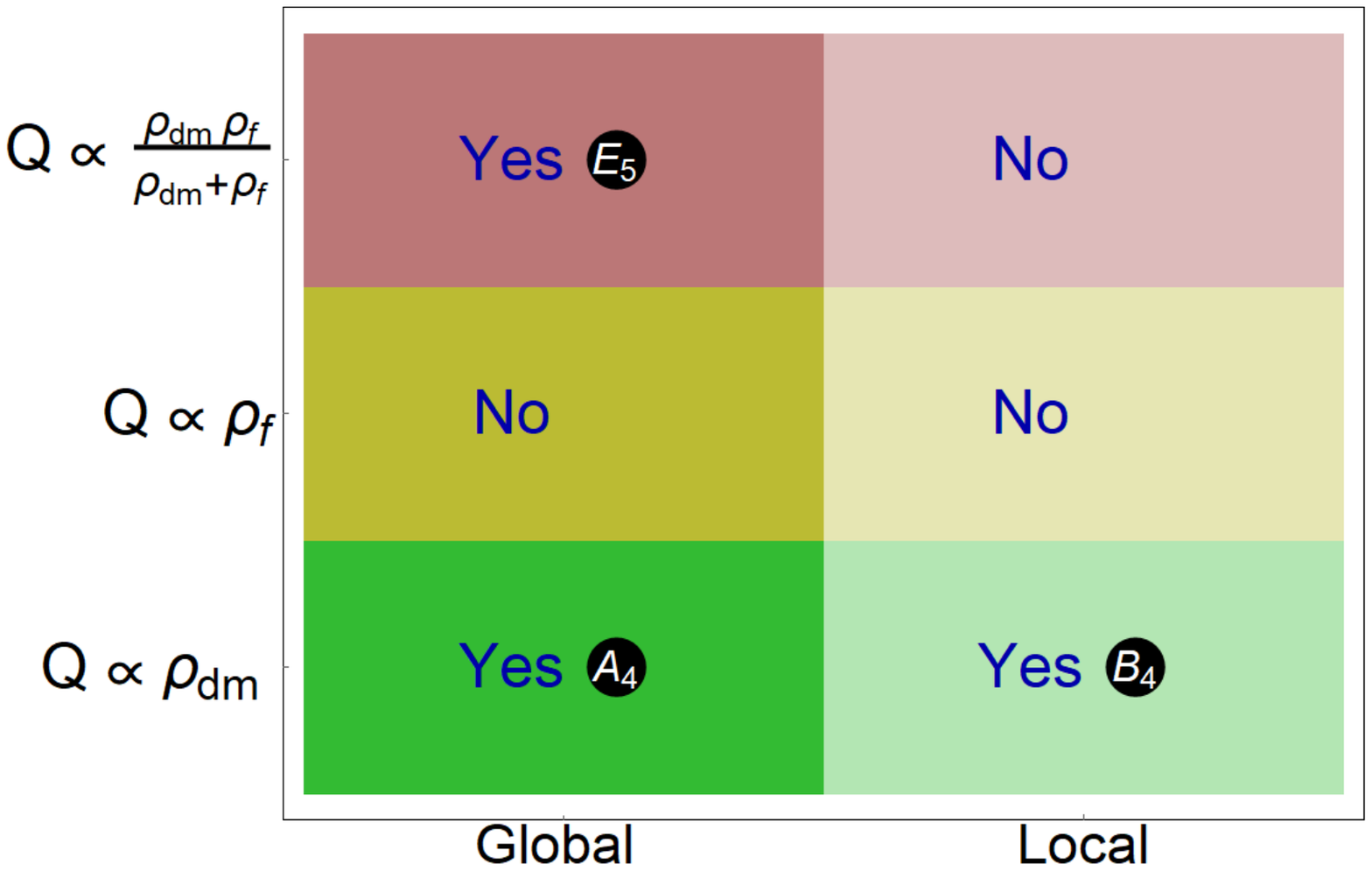}
    \caption{Summary of accelerating scaling attractors identified in this 
study. Each row corresponds to a particular energy density (green, yellow, or 
pink), while each column represents the type of interaction: global rate (dark 
shading) or local rate (light shading). }
    \label{fig7}
\end{figure}

The dynamical analysis of all interaction models (Tables 
\ref{tab:model1-positive-w}-\ref{tab:model3-positive-w}) clearly shows that 
the 
existence of a late-time accelerating scaling attractor, i.e. a stable 
critical point with $q<0$ and $0<\Omega_{\rm dm}<1$, is extremely sensitive to 
both the functional dependence of the interaction rate $Q$ and the direction of 
energy transfer between the two fluids.
 The 
summary Fig. \ref{fig7} highlights these 
findings very clearly, offering a comprehensive overview of how different 
interaction prescriptions shape the late-time dynamics.
A systematic comparison 
of the models reveals several general conditions under which such attractors 
can 
arise. 

A first important condition is that the interaction must depend explicitly on 
the DM density. Accelerating scaling solutions are found only in the cases 
where 
the interaction takes the form $Q=\mu H\rho_{\rm dm}$ or $Q=\Gamma\rho_{\rm 
dm}$. In both of these models the dynamical system admits stable coexistence 
points (A$_4$ and B$_4$), where the universe undergoes late-time accelerated 
expansion with $q=-1$ and with the two fluids sharing the total energy density, 
$0<\Omega_{\rm dm}<1$ and $\Omega_{\rm dm}+\Omega_f=1$. These models therefore 
provide genuine accelerating scaling attractors.

A second requirement concerns the direction of energy transfer. In the models 
that allow scaling attractors (A and B), the compact phase space $\mathbb{D}$ 
is 
positively invariant only when $\mu>0$ or $\beta>0$, implying that energy flows 
from the matter-creating DM component into the second fluid. If the direction 
of 
energy flow is reversed, namely $\mu<0$ or $\beta<0$, the phase space loses 
positive invariance and the accelerating scaling attractors disappear. This 
behaviour is evident in the phase portraits of Models~A and B.

By contrast, interactions proportional to the energy density of the second 
fluid 
cannot produce accelerating scaling attractors. In both the global and local 
cases, $Q=\xi H\rho_f$ and $Q=\widetilde{\Gamma}\rho_f$, the late-time 
attractors (C$_3$, D$_3$) are always fully DM-dominated, even though they 
correspond to accelerated expansion with $q=-1$. All coexistence critical 
points 
in these models are either saddles or unstable, and thus none of them can serve 
as scaling attractors.

Mixed interactions display a more nuanced behaviour. In the global mixed case, 
namely
\[
Q=\delta H \frac{\rho_{\rm dm}\rho_f}{\rho_{\rm dm}+\rho_f},
\]
the critical point E$_5$ becomes a stable accelerating scaling solution 
whenever 
$\delta>0$ and $\delta\geq 3(1+w)$, as shown in Table 
\ref{tab:model3-positive-w} and Fig. \ref{fig5}. Under 
these conditions the two fluids coexist with constant fractional densities and 
$q=-1$. However, the local mixed interaction, i.e.
\[
Q=\overline{\Gamma}\frac{\rho_{\rm dm}\rho_f}{\rho_{\rm dm}+\rho_f},
\]
does not admit any stable accelerating coexistence point: the accelerating 
coexistence solutions (F$_5$, F$_6$) are saddles or unstable, and the only 
stable accelerating solutions (F$_3$, F$_4$) are fully DM-dominated. 
Consequently, no accelerating scaling attractor is realised in this case.

In summary, these results show that accelerating scaling attractors exist 
if 
and only if three requirements are simultaneously satisfied. Firstly, the 
interaction rate must involve the DM density or the global mixed density, 
namely $Q\propto\rho_{\rm dm}$ or $Q\propto H\,\rho_{\rm dm}\rho_f/(\rho_{\rm 
dm}+\rho_f)$. Secondly, energy transfer must occur from the matter-creating DM 
sector to the second fluid, so that $Q>0$. Thirdly, the matter-creation rate 
$\Psi_0$ must remain positive (equivalently $\alpha>0$), ensuring a negative 
creation pressure capable of cooperating with the interaction in stabilising 
the 
accelerating branch. These features, summarised in Fig.~\ref{fig7}, demonstrate 
that only 
the combined effect of matter creation and DM-based interactions can produce 
stable accelerating scaling cosmologies, a phenomenon absent in non-interacting 
matter-creation scenarios.

\subsection{Role of energy-transfer direction and matter creation}

The dynamical analysis makes it clear that the late-time evolution of the 
system 
is determined not only by the specific functional dependence of the interaction 
rate $Q$, but also it is critically determined  by the direction of energy 
transfer between 
the two fluids and by the presence of a positive matter-creation rate. These 
two 
ingredients together select whether the universe ends in an accelerating 
scaling 
state, in a purely accelerating DM-dominated phase, or in a decelerating 
configuration. The comparison among all interaction classes reveals several 
recurrent mechanisms.

A first key result is that positive energy transfer, $Q>0$, is necessary for 
the 
emergence of stable accelerating coexistence solutions. When energy flows from 
the matter-creating DM component to the second fluid, stable late-time 
accelerating points appear in Models~A, B, and E, corresponding respectively to 
A$_4$, B$_4$, and E$_5$. In such cases the interaction acts as a regulating 
mechanism, maintaining a fixed ratio $\Omega_{\rm dm}/\Omega_f$ and preventing 
either component from fully dominating. The resulting critical points 
constitute 
genuine accelerating scaling attractors, with $q<0$ and $0<\Omega_{\rm dm}<1$, 
stabilised by the combined effects of energy depletion from DM and the 
balancing 
contribution of matter creation.

The opposite direction of energy flow, $Q<0$, immediately destabilises the 
phase 
space and removes all accelerating scaling attractors. When energy is 
transferred from the second fluid into DM, the compact phase space $\mathbb{D}$ 
ceases to be positively invariant in Models~A and B; trajectories approach the 
boundaries and no stable coexistence solution survives. Physically, reverse 
energy transfer reinforces the continuous growth of the DM sector caused by 
matter creation, thereby driving the system irreversibly toward DM domination 
and suppressing the possibility of stabilising a mixed accelerating state.

A similar suppression of scaling behaviour occurs in all models where the 
interaction rate depends on the density of the second fluid. In Models~C and D 
the interaction strength is proportional to $\rho_f$, making the 
energy-transfer 
term too weak to counteract the monotonic growth of the DM density induced by 
matter creation. As a consequence, the late-time attractors C$_3$ and D$_3$ are 
always fully DM-dominated accelerated universes. All coexistence points in 
these 
models are saddles or unstable, demonstrating that interactions governed by 
$\rho_f$ are dynamically subdominant compared with the persistent injection of 
DM.

Mixed interactions show a split behaviour that depends on how strongly the DM 
sector controls the transfer rate. The global mixed interaction of Model~E 
includes the symmetric combination $\rho_{\rm dm}\rho_f/(\rho_{\rm dm}+\rho_f)$ 
modulated by the Hubble rate; this structure is sufficiently robust to produce 
the stable accelerating scaling attractor E$_5$ when $\delta>0$. In contrast, 
the local mixed interaction of Model~F lacks this stabilising component: its 
transfer terms are too weak to offset matter creation, and the only stable 
accelerating fixed points (F$_3$, F$_4$) are purely DM-dominated. Thus no 
accelerating scaling attractor is produced in Model~F.

Finally, the presence of a positive matter-creation rate, $\Psi_0>0$, plays a 
decisive role in shaping the dynamics. Matter creation introduces a negative 
creation pressure,
\[
p_c=-\frac{\Psi_0}{3H}\rho_{\rm dm},
\]
which naturally drives accelerated expansion but simultaneously enhances the DM 
density. Matter creation alone therefore pushes the system toward DM 
domination. 
A scaling attractor can emerge only if the interaction efficiently transfers a 
sufficiently large fraction of the newly created DM into the second fluid. This 
explains why only the interaction models explicitly depending on $\rho_{\rm 
dm}$ 
(Models~A, B, and E) generate stable accelerating scaling solutions, whereas 
the 
interaction prescriptions controlled by $\rho_f$ or by local mixed terms 
(Models~C, D, and F) do not.

Altogether, these results show that the coexistence of cosmic acceleration and 
stable scaling behaviour is realised only when matter creation is supported by 
a 
DM-driven forward energy transfer. This cooperative mechanism counteracts the 
continuous production of DM, stabilises the ratio of the energy densities, and 
produces an accelerating attractor without the need to invoke a dark-energy 
component.

\subsection{Unified comparison of all models}

The dynamical analysis carried out in the previous section shows that the 
six interaction prescriptions fall naturally into two broad categories, each 
characterised by a distinct set of late-time behaviours. Across all models the 
phase space exhibits only three possible asymptotic outcomes: a stable 
accelerating scaling state with $0<\Omega_{\rm dm}<1$, a stable accelerating 
DM-dominated attractor, or unstable and saddle decelerating configurations. The 
key discriminator among the six models is therefore the presence or absence of 
the first class, namely the accelerating scaling solutions.

Models~A, B, and E admit stable accelerating coexistence states (A$_4$, B$_4$, 
and E$_5$), provided that energy is transferred from the matter-creating DM 
component to the second fluid. Under this condition the interaction 
counterbalances the continuous growth of the DM sector induced by matter 
creation, allowing the ratio $\Omega_{\rm dm}/\Omega_f$ to settle to a constant 
value. These are the only scenarios in which matter creation, supplemented by a 
suitable interaction, can reproduce an effective dark-energy behaviour while 
preserving a non-trivial mixture of cosmic components.

By contrast, Models~C, D, and F do not produce a stable accelerating 
coexistence 
state for any admissible parameter choice. Their corresponding late-time 
attractors (C$_3$, D$_3$, and F$_3$-F$_4$) are invariably DM-dominated, 
irrespective of the direction of energy transfer. The common feature of these 
models is that their interaction terms either depend on the density of the 
second fluid or involve a local mixed rate that scales too weakly to compete 
with the persistent increase of $\rho_{\rm dm}$ generated by matter creation. 
As 
a result, the DM sector inevitably dominates before any mixed accelerating 
state 
can stabilise.

The underlying mechanism behind this grouping is straightforward. Because 
matter 
creation continuously amplifies the DM density, any viable accelerating scaling 
attractor requires an interaction strong enough to remove energy from the DM 
sector at a rate comparable to the creation rate itself. Only interactions 
proportional to $\rho_{\rm dm}$, or to a symmetrised density factor dominated 
by 
$\rho_{\rm dm}$, satisfy this condition. Interactions controlled by $\rho_f$ or 
by local mixed combinations remain dynamically subdominant and are therefore 
unable to prevent the system from evolving toward a DM-dominated accelerating 
universe.

In summary, the six models organise themselves into a clear hierarchy. 
Models~A, 
B, and E can generate stable accelerating scaling attractors when $Q>0$ and are 
therefore capable of mimicking late-time cosmic acceleration without invoking a 
dark-energy component. Models~C, D, and F, on the other hand, lack such 
attractors; their late-time dynamics is always DM-dominated, and acceleration 
arises solely from the creation pressure. This global classification highlights 
the decisive role of DM-based interactions in determining whether matter 
creation alone can reproduce an accelerating universe with a dynamically 
balanced cosmic composition.

\section{Summary and Conclusions}
\label{sec-summary}

In this work we have explored a general two-fluid cosmology in which a 
pressureless dark-matter component experiences constant-rate matter creation 
while interacting with a barotropic fluid  with $w \geq 0$. Six 
well-studied interaction 
prescriptions were analysed, covering global and local forms and including 
cases 
where the energy transfer depends on $\rho_{\rm dm}$, $\rho_f$, or a mixed 
density combination. Through a detailed dynamical-systems analysis over a 
compact phase space, we identified all critical points, their stability 
properties, and their cosmological interpretation. This allowed us to determine 
which interaction forms are capable of generating late-time accelerated 
expansion and under what conditions.

Our results reveal a sharp division among the six interaction models. Whenever 
the interaction rate depends explicitly on the dark-matter density or on a 
global mixed-density term, stable accelerating scaling attractors emerge at 
late 
times, provided that energy flows from the matter-creating dark matter to the 
second fluid. These scaling solutions exhibit accelerated expansion while 
maintaining a non-trivial balance between the two fluids, and they arise 
without 
requiring any dark-energy-like equation of state. In contrast, interactions 
governed by the density of the second fluid, as well as local mixed 
interactions, never produce such attractors: in these models the universe 
invariably evolves toward a dark-matter-dominated accelerating phase, since the 
continuous creation of matter overwhelms any subdominant interaction effects. 
This unified behaviour, summarised in our phase-space diagrams and tables, 
highlights a consistent dynamical mechanism across all models.

\section*{Acknowledgments}
We thank the referee for some essential comments that became useful to improve the manuscript.
SH acknowledges the financial support from the University Grants Commission 
(UGC), Govt. of India (NTA Ref. No: 201610019097). JdH is supported by the 
Spanish grants PID2021-123903NB-I00 and RED2022-134784-T funded by 
MCIN/AEI/10.13039/501100011033 and by ERDF ``A way of making Europe''.  SP and 
TS acknowledge the partial support from the Department of Science and 
Technology (DST), Govt. of India under the Scheme   ``Fund for Improvement of 
S\&T Infrastructure (FIST)'' (File No. SR/FST/MS-I/2019/41). ENS acknowledges 
the contribution of the LISA CosWG, and of   COST Actions  CA18108  "Quantum 
Gravity Phenomenology in the multi-messenger approach''  and  CA21136" 
Addressing observational tensions in cosmology with systematics and fundamental 
physics (CosmoVerse)''.

\appendix

\section{Tables}
\label{sec-appendix}

In this section we show the Tables~\ref{tab:model1-positive-w} $-$ \ref{tab:model3-positive-w} presenting the critical points, their existences, stabilities and the cosmological parameters evaluated at those points. 

\renewcommand{\thetable}{A\arabic{table}} 
\begin{table*}[]
    \centering
    \resizebox{1.0\textwidth}{!}{%
    \begin{tabular}{|c|c|c|c|c|c|}
    
    \hline\hline
    \multicolumn{6}{|c|}{Model: $Q=\mu H \rho_{\rm dm}$}\\
    \hline
      Critical Points & Existence & Eigenvalues & Stability & $\Omega_{dm}$ & 
Acceleration    \\ \hline   
     
        $A_0\left(0,0\right)$     & Always     & 
$\left(\frac{3}{2}(1+w),~3w-\mu\right)$ & Unstable if $\mu<3w$; Saddle if 
$\mu\geq 3w$ 
       & $0$ & No: $q=\frac{1}{2}(1+3w)$     \\

      $A_1\left(0,1\right)$ & Always & $(\alpha,~0)$ &  Non-hyperbolic Saddle   
& $0$ & Undetermined     \\

      $A_2\left(1,1\right)$ & Always  & 
$\left(-\alpha,~\frac{\alpha}{2}\right)$ 
& Saddle     &    $1$ & Yes: $q\longrightarrow-\infty$        \\

       $A_3\left(1-\frac{\mu}{3w},0\right)$    & $\mu\leq3w$, $w\neq 0$      & 
$\left(\mu-3w,~\frac{1}{2}(\mu+3)\right)$ & Saddle   & $1-\frac{\mu}{3w}$ & No: 
$\frac{1}{2}(1+\mu)$       \\ 
     
      $A_4\left(\frac{3(1+w)}{3(1+w)+\mu},\frac{3+\mu}{3+\mu+\alpha}\right)$    
 
 &  Always     & $\left(\lambda_1,~\lambda_2\right)$     &  {\bf Stable}     &  
$\frac{3(1+w)}{3(1+w)+\mu}$       & Yes: $q=-1$       \\  \hline\hline
     
    \multicolumn{6}{|c|}{Model: $Q=\Gamma \rho_{\rm dm}$}\\
    \hline
      Critical Points & Existence & Eigenvalues & Stability & $\Omega_{dm}$ & 
Acceleration   \\ \hline

      $B_0\left(0,0\right)$     & Always     & 
$\left(3w,~\frac{3}{2}(1+w)\right)$ & Unstable  & $0$ & No: 
$q=\frac{1}{2}(1+3w)$     \\

      $B_1\left(0,1\right)$ & Always & $(\alpha-\beta,~0)$ &  Saddle if 
$\alpha>\beta$; {\bf Stable} if $\alpha\leq\beta$   & $0$ & Undetermined     \\

      $B_2\left(1,0\right)$ & Always  & $\left(-3w,~\frac{3}{2}\right)$ & 
Saddle 
if $w>0$; Unstable if $w=0$     &    $1$ & No: $q=\frac{1}{2}$        \\

      $B_3\left(1-\frac{\beta}{\alpha},1\right)$    & $\alpha \geq \beta$      
& 
$\left(\beta-\alpha,~\frac{\alpha-\beta}{2}\right)$ & Saddle if $\alpha>\beta$; 
{\bf Stable} if $\alpha=\beta$   & $1-\frac{\beta}{\alpha}$ & Yes: 
$q\longrightarrow-\infty$       \\

$B_4\left(\frac{(1+w)(\alpha-\beta)}{(1+w)(\alpha-\beta)+\beta},\frac{3}{
3+\alpha-\beta}\right)$      &  $\alpha \geq \beta$    & 
$\left(\psi_1,~\psi_2\right)$     &  {\bf Stable}     &  
$\frac{(1+w)(\alpha-\beta)}{(1+w)(\alpha-\beta)+\beta}$       & Yes: $q=-1$     
 
 \\  
      
      $B_5\left(x_c,0\right)$ & $0 \leq x_c \leq 1$, $w=0$ & 
$\left(\frac{3}{2},~0\right)$ &  Unstable  & $x_c$ & No: $q=\frac{1}{2}$     \\ 
  \hline\hline
  
  \end{tabular}%
    }
    \caption{Summary table of   model A: $Q=\mu H \rho_{\rm dm}$ and model B: 
$Q=\Gamma\rho_{\rm dm}$, assuming $w\geq0$, with the properties of the critical 
points and the corresponding values of the 
cosmological parameters. Here, 
$\lambda_{1,2}=\frac{1}{C}\left(A\pm \sqrt{B}\right)$ where $A,~B$ and $C$ are 
defined as $A=-9\alpha(1+w)\left(6w^2+5w(3+\mu)+(3+\mu)^2\right)$, 
$B=3\alpha^2(1+w)(3+3w+\mu)^2\left(108w^3+36w^2(6+\mu)-(3+\mu)^2(8\mu-3)-3w(7\mu
^2+6\mu-45)\right)$ and $C=4(3+3w+\mu)^2(3+\alpha+\mu)$, respectively. 
Moreover, 
$\psi_{1,2}=\frac{\alpha-\beta}{D}\left(E\pm \sqrt{F}\right)$ where $D,~E$ and 
$F$ are defined as $D=4(3+\alpha-\beta)(\alpha+w\alpha-w\beta)^2$, 
$E=-3(1+w)\left((3+5w+2w^2)\alpha^2-w(5+4w)\alpha\beta+2w^2\beta^2\right)$ and 
$F=9(1+w)(\alpha+w\alpha-w\beta)^2\left((1+w)
(\alpha+2w\alpha)^2-4w(2w^2+w-1)\alpha\beta+4(w-1)w^2\beta^2\right)$, 
respectively. }
    \label{tab:model1-positive-w}
\end{table*}
\begin{table*}[]
    \centering
    \resizebox{0.95\textwidth}{!}{%
    \begin{tabular}{|c|c|c|c|c|c|}
    
    \hline\hline
    \multicolumn{6}{|c|}{Model: $Q=\xi H \rho_{f}$}\\
    \hline
      Critical Points & Existence & Eigenvalues & Stability & $\Omega_{dm}$ & 
Acceleration    \\ \hline   
     
     $C_0\left(0,1\right)$     & Always     & $\left(\alpha,~0\right)$ & 
Non-hyperbolic Saddle  
       & $0$ & Undetermined    \\

     $C_1\left(1,0\right)$ & Always & $\left(\xi-3w,~\frac{3}{2}\right)$ &  
Saddle   & $1$ & No: $q=\frac{1}{2}$     \\

      $C_2\left(1,1\right)$ & Always  & 
$\left(-\alpha,~\frac{\alpha}{2}\right)$ 
& Saddle     &    $1$ & Yes: $q\longrightarrow-\infty$        \\

      $C_3\left(1,\frac{3}{3+\alpha}\right)$    & Always      & 
$\left(\frac{\alpha(\xi-3w-3)}{3+\alpha},~-\frac{3\alpha}{2(3+\alpha)}\right)$ 
& 
{\bf Stable}   & $1$ & Yes: $q=-1$       \\  \hline  \hline

    \multicolumn{6}{|c|}{Model: $ Q = \widetilde{\Gamma}  \rho_{f} $}\\
    \hline
     Critical Points & Existence & Eigenvalues & Stability & $\Omega_{dm}$ & 
Acceleration    \\ \hline
     
     $D_0\left(0,0\right)$     & Always     & 
$\left(3w,~\frac{3}{2}(1+w)\right)$ & Unstable  & $0$ & No: 
$q=\frac{1}{2}(1+3w)>0$     \\

     $D_1\left(1,0\right)$ & Always & $\left(-3w,~\frac{3}{2}\right)$ &  Saddle 
if $w>0$   & $1$ & No: $q=\frac{1}{2}$    \\

    $D_2\left(1,1\right)$ & Always  & 
$\left(\gamma-\alpha,~\frac{\alpha}{2}\right)$ & Saddle     &    $1$ & Yes: 
$q\longrightarrow-\infty$        \\

   $D_3\left(1,\frac{3}{3+\alpha}\right)$    & Always      & 
$\left(\frac{3(\gamma-\alpha-w\alpha)}{3+\alpha},~-\frac{3\alpha}{2(3+\alpha)}
\right)$ & {\bf Stable}   & $1$ & Yes: $q=-1$       \\   

      $D_4\left(x_c,0\right)$      & $0\leq x_c \leq 1$, $w=0$    & 
$\left(0,~\frac{3}{2}\right)$     &  Unstable     &  $x_c$       & No: 
$q=\frac{1}{2}$       \\  \hline\hline
      \end{tabular}%
    }
    \caption{Summary table of the models $Q=\mu H \rho_{f}$ and 
$Q=\widetilde{\Gamma}\rho_{f}$, considering second fluid equation of state as 
non negative $(w\geq0)$, where properties of the critical points and values of 
the cosmological parameters at these points are highlighted. }
    \label{tab:model2-positive-w}
\end{table*}
\begin{table*}[]
    \centering
    \resizebox{1.0\textwidth}{!}{%
    \begin{tabular}{|c|c|c|c|c|c|}
    
    \hline\hline
    \multicolumn{6}{|c|}{Model: $ Q = \delta H \rho_{\rm dm} \rho_{f}(\rho_{\rm 
dm} + \rho_{f})^{-1} $}\\
    \hline
      Critical Points & Existence & Eigenvalues & Stability & $\Omega_{dm}$ & 
Acceleration    \\ \hline
     
     $E_0\left(0,0\right)$     & Always     & 
$\left(3w-\delta,~\frac{3}{2}(1+w)\right)$ & Saddle if $\delta>3w$; Unstable if 
$\delta\leq 3w$  & $0$ & No: $q=\frac{1}{2}(1+3w)$     \\

     $E_1\left(1,0\right)$ & Always & $\left(\delta-3w,~\frac{3}{2}\right)$ &  
Saddle if $\delta<3w$; Unstable if $\delta\geq 3w$   & $1$ & No: 
$q=\frac{1}{2}$ 
   \\

     $E_2\left(0,1\right)$ & Always  & $\left(\alpha,~0\right)$ &  
Non-hyperbolic Saddle   &    $0$ & Undetermined       \\

     $E_3\left(1,1\right)$ & Always  & $\left(-\alpha,~\frac{\alpha}{2}\right)$ 
& Saddle  &    $1$ & Yes: $q\longrightarrow-\infty$        \\

    $E_4\left(1,\frac{3}{3+\alpha}\right)$    & Always      & 
$\left(\frac{\alpha(\delta-3w-3)}{3+\alpha},~-\frac{3\alpha}{2(3+\alpha)}
\right)$ & {\bf Stable} if $\delta\leq 3w+3$; Saddle if $\delta>3w+3$   & $1$ & 
Yes: $q=-1$       \\  

     $E_5\left(\frac{3(1+w)}{\delta},\frac{\delta-3w}{\delta-3w+\alpha}\right)$ 
 
  & $\delta\geq 3(1+w)$   & $\left(\Psi_1,~\Psi_2 \right)$ & {\bf Stable}   & 
$\frac{3(1+w)}{\delta}$ & Yes: $q=-1$       \\ 
     
     $E_6\left(x_c,0\right)$ & $\delta=3w$, $w\neq 0$,  & 
$\left(0,~\frac{3}{2}\left(1+w\left(1-x_c\right)\right)\right)$ & Unstable  &   
 
$x_c$ & No: $q>0$        \\ 

    &   $0\leq x_c \leq 1$    &       &       &         &         \\ 
\hline\hline
    
     \multicolumn{6}{|c|}{Model: $ Q = \overline{\Gamma} \rho_{\rm dm} \rho_{f} 
(\rho_{\rm dm} + \rho_{f})^{-1} $}\\
    \hline
     Critical Points & Existence & Eigenvalues & Stability & $\Omega_{dm}$ & 
Acceleration    \\ \hline

     $F_0\left(0,0\right)$     & Always     & 
$\left(3w,~\frac{3}{2}(1+w)\right)$ & Unstable  & $0$ & No: 
$q=\frac{1}{2}(1+3w)$     \\

     $F_1\left(1,0\right)$ & Always & $\left(-3w,~\frac{3}{2}\right)$ &  Saddle 
 
& $1$ & No: $q=\frac{1}{2}$    \\

     $F_2\left(0,1\right)$ & Always  & $\left(\alpha-\nu,~0\right)$ &  Saddle 
if 
$\nu<\alpha$; {\bf Stable} if $\nu>\alpha$  &  $0$ & Undetermined       \\

     $F_3\left(1,1\right)$ & Always  & 
$\left(\nu-\alpha,~\frac{\alpha}{2}\right)$ & Saddle if $\nu<\alpha$; Unstable 
if $\nu>\alpha$  &    $1$ & Yes: $q\longrightarrow-\infty$        \\

     $F_4\left(1,\frac{3}{3+\alpha}\right)$    &  $\alpha>0$ & 
$\left(\frac{3(\nu-\alpha(1+w))}{3+\alpha},~-\frac{3\alpha}{2(3+\alpha)}
\right)$ 
& {\bf Stable} if $\nu\leq \alpha(1+w)$; Saddle if $\nu>\alpha(1+w)$  & $1$ & 
Yes: $q=-1$       \\  

     $F_5\left(\frac{(1+w)(\nu-\alpha)}{\nu w},\frac{3w}{3w+\nu-\alpha}\right)$ 
 
  & $w\neq 0$, $\nu\geq\alpha$,    & $\left(\Phi_1,~\Phi_2 \right)$ & Saddle   
& 
$\frac{(1+w)(\nu-\alpha)}{\nu w}$ & Yes: $q=-1$       \\ 
     &  $\alpha(1+w)-\nu\geq 0$    &      &       &       &     \\
     
     $F_6\left(x_c,1\right)$ & $0\leq x_c \leq 1$, $\alpha=\nu$  & 
$\left(0,~\frac{\nu x_c}{2}\right)$ & Unstable  & $x_c$ & Yes: 
$q\longrightarrow 
-\infty$        \\   
     
     $F_7\left(x_c,0\right)$ & $0\leq x_c \leq 1$, $w=0$  & 
$\left(0,~\frac{3}{2}\right)$ & Unstable  & $x_c$ & No: $q=\frac{1}{2}$        
\\  \hline\hline
     \end{tabular}%
    }
    \caption{Summary table of the models $ Q = \delta H \rho_{\rm dm} 
\rho_{f}(\rho_{\rm dm} + \rho_{f})^{-1} $ and $ Q = \overline{\Gamma} \rho_{\rm 
dm} \rho_{f}(\rho_{\rm dm} + \rho_{f})^{-1} $, considering second fluid 
equation 
of state as non-negative $(w\geq0)$, where properties of the critical points 
and 
values of the cosmological parameters at these points are highlighted. Here, 
$\Psi_{1,2}=\frac{1}{O}\left(M\pm \sqrt{N}\right)$ where $O,~M$ and $N$ are 
defined as $O=4\delta(3w-\alpha-\delta)$, $M=-3\alpha(1+w)(3w-\delta)$ and 
$N=3\alpha^2(1+w)(3w-\delta)\left(9w+9w^2-27\delta-27w\delta+8\delta^2\right)$, 
respectively. Again, $\Phi_{1,2}=\frac{\alpha-\nu}{R}\left(S\pm 
\sqrt{T}\right)$ 
where $R,~S$ and $T$ are defined as $R=4\nu(3w-\alpha+\nu)$, $S=3\alpha(1+w)$ 
and $T=9(1+w)\left((1+w)\alpha^2+8(1+w)\alpha\nu-8\nu^2\right)$, respectively. }
    \label{tab:model3-positive-w}
\end{table*}

\bibliographystyle{elsarticle-num}
\bibliography{biblio}

\end{document}